\documentclass[10pt,journal,a4paper,twoside,twocolumn]{IEEEtran}
\usepackage{cite}
\usepackage{float}
\usepackage{stfloats}
\usepackage{fancyhdr}
\usepackage{color}
\usepackage[inline]{enumitem}
\usepackage{amsmath,amsfonts,amssymb}
\usepackage{graphicx}

\usepackage{algorithm}
\usepackage{algorithmic}
\usepackage{epsfig}
\usepackage{epstopdf}
\usepackage[caption=false]{subfig}
\usepackage{bm}
\usepackage{array}
\usepackage{bbding}

\usepackage{bbm}

\newtheorem{Rem}{Remark}

\begin{document}
	\title{Bedrock Models in Communication and Sensing: Advancing Generalization, Transferability, and Performance
	}
	
	\author{\IEEEauthorblockN{
		Cheng Luo, \IEEEmembership{Graduate Student Member, IEEE}, Luping Xiang, \IEEEmembership{Senior Member, IEEE}, Jie Hu, \IEEEmembership{Senior Member, IEEE} and Kun Yang, \IEEEmembership{Fellow, IEEE}
		}
			\\

		\thanks{Cheng Luo and Jie Hu are with the School of Information and Communication Engineering, University of Electronic Science and Technology of China, Chengdu, 611731, China, email: chengluo@std.uestc.edu.cn; hujie@uestc.edu.cn.}
		\thanks{Luping Xiang is with the State Key Laboratory of Novel Software Technology, Nanjing University, Nanjing, 210008, China, and School of Intelligent Software and Engineering, Nanjing University (Suzhou Campus), Suzhou, email: luping.xiang@nju.edu.cn.}
		\thanks{Kun Yang is with the School of Computer Science and Electronic Engineering, University of Essex, Colchester CO4 3SQ, U.K., email: kunyang@essex.ac.uk.}
		\thanks{(Corresponding author: Luping Xiang.)}
		
	}

	\maketitle
	
	\thispagestyle{fancy} 
	\lhead{} 
	\chead{} 
	\rhead{} 
	\lfoot{} 
	\cfoot{} 
	\rfoot{\thepage} 
	\renewcommand{\headrulewidth}{0pt} 
	\renewcommand{\footrulewidth}{0pt} 
	\pagestyle{fancy}

    \rfoot{\thepage} 
	\begin{abstract}
	Deep learning (DL) has emerged as a powerful tool for addressing the intricate challenges inherent in communication and sensing systems, significantly enhancing the intelligence of future sixth-generation (6G) networks. A substantial body of research has highlighted the promise of DL-based techniques in these domains. However, in addition to improving accuracy, new challenges must be addressed regarding the generalization and transferability of DL-based systems. To tackle these issues, this paper introduces a series of mathematically grounded and modularized models, referred to as bedrock models, specifically designed for integration into both communication and sensing systems. Due to their modular architecture, these models can be seamlessly incorporated into existing communication and sensing frameworks. For communication systems, the proposed models demonstrate substantial performance improvements while also exhibit strong transferability, enabling direct parameter sharing across different tasks, which greatly facilitates practical deployment. In sensing applications, the integration of the bedrock models into existing systems results in superior performance, reducing delay and Doppler estimation errors by an order of magnitude compared to traditional methods. Additionally, a pre-equalization strategy based on the bedrock models is proposed for the transmitter. By leveraging sensing information, the transmitted communication signal is dynamically adjusted without altering the communication model pre-trained in AWGN channels. This adaptation enables the system to effectively cope with doubly dispersive channels, restoring the received signal to an AWGN-like condition and achieving near-optimal performance. Simulation results substantiate the effectiveness and transferability of the proposed bedrock models, underscoring their potential to advance both communication and sensing systems.
	\end{abstract}

	\begin{IEEEkeywords}
		Deep learning, communications, sensing, sensing-aided communications, transferability.
	\end{IEEEkeywords}

	\section{Introduction}

	\IEEEPARstart{T}{he} successful deployment and widespread adoption of fifth-generation (5G) networks have catalyzed preliminary discussions surrounding the design of sixth-generation (6G) communication systems. This shift is largely driven by the need to accommodate emerging applications such as augmented/virtual reality, autonomous driving, high-precision manufacturing, and the rapidly growing Internet of Everything (IoE) \cite{tutorial2, Com5G26G, 10783105, 9349624}. These advancements, however, pose significant challenges to the existing communication frameworks, including issues related to spectrum sensing, energy efficiency, low-latency transmission, and high data throughput\cite{JilinWang_OFDMsensing, DD_sep, RISS}.

	In addition, as future communication networks are anticipated to become increasingly intricate and adaptive, the traditional modular-layered network architectures may prove inadequate for optimal end-to-end (E2E) communication. Furthermore, established model-based methods and algorithms may no longer be sufficient, particularly due to the lack of explicit mathematical models and the excessive complexity associated with real-time, cross-layer optimizations, which could result in unmanageable system overhead.

	In response to these challenges, deep learning (DL)-based intelligent wireless communication networks, characterized by their capabilities for self-organization, self-configuration, and self-recovery, have emerged as a promising approach \cite{tutorial1, tutorial3, tutorial4, LLM4Comm, 8743390}. These networks are particularly well-suited for addressing complex tasks that involve cross-layer joint optimization or are too unpredictable to model in advance, representing a significant step toward the future of communication technologies\cite{9206115}.

	Recent studies have explored various DL-driven components for the physical layer of wireless communication systems, including constellation mapping \cite{constellationdesign2, OshePhysical}, pilot and waveform design \cite{pilotdesign, waveform2}, channel coding \cite{coding1, coding2}, E2E transceiver design \cite{constellationNorm, CP_zhao}, and radio frequency (RF) fingerprint classification \cite{RF_recog}. Notably, \cite{constellationdesign2} and \cite{OshePhysical} propose autoencoder (AE)-based communication systems with trainable constellation mapping, enabling joint optimization of both constellation shaping and labeling at the transmitter. In the area of pilot and waveform design, \cite{pilotdesign} introduces a neural network (NN)-based scheme for joint pilot design and downlink channel estimation, which effectively reduces pilot transmission overhead by pruning less significant neurons from the NN layers. Additionally, \cite{waveform2} develops an orthogonal frequency-division multiplexing (OFDM) waveform for the joint optimization of transmit and receive filters, addressing constraints on adjacent channel leakage ratio (ACLR) and peak-to-average power ratio (PAPR), thus achieving a balanced trade-off between information rate, ACLR, and PAPR. On the coding front, \cite{coding1, coding2} frame Polar coding/decoding and modulation/demodulation as an AE, demonstrating superior performance in both wireless communication and energy transfer. Furthermore, DL can be leveraged to develop novel E2E communication architectures, such as in \cite{constellationNorm}, where an E2E transceiver is used for pilot selection and superimposition to reduce pilot overhead, and for constellation mapping training to enhance throughput in doubly dispersive channels. Similarly, \cite{CP_zhao} proposes a transceiver and equalization system based on complex convolution, utilizing the cyclic prefix (CP) to improve communication performance.

	Sensing performance and integrated sensing and communication (ISAC) systems, which are critical for 6G networks, have also greatly benefited from DL methodologies \cite{MUSIC_enhanced, Deep-MUSIC1, Deep-MUSIC2, waveform_ISAC, ISAC_NET}. The widely used multiple signal classification (MUSIC) algorithm, applied in frequency estimation, has been enhanced using DL techniques \cite{MUSIC_enhanced, Deep-MUSIC1, Deep-MUSIC2}. Specifically, \cite{MUSIC_enhanced} introduces a network that leverages sampled received signals, rather than covariance matrices, as input to extract features, thereby addressing imperfections such as inconsistent gains/phases and nonlinear amplifier effects in direction-of-arrival (DOA) estimation. In parallel, \cite{Deep-MUSIC1, Deep-MUSIC2} processes the empirical covariance matrix (ECM), reconstructs the noise subspace, and outputs continuous, grid-free estimates via neural networks (NNs), resulting in superior performance. In the domain of ISAC systems, \cite{waveform_ISAC} proposes a novel OFDM waveform that replaces the CP with a flexible guard interval, reducing CP overhead and enhancing data rates. Two NN-based networks are designed for data extraction and parameter estimation, respectively. Additionally, \cite{ISAC_NET} introduces an ISAC signal processing optimization scheme that jointly processes pilot and data signals with an iterative optimization algorithm, improving both sensing accuracy and communication demodulation. A summary of related works is presented in Table \ref{table:relatedwork}.

	\begin{table*}[]
		\centering
		\caption{Relevant works on DL-based communication and sensing systems.}
		\begin{tabular}{m{3 cm}<{\centering}|m{1.2 cm}<{\centering}|m{1.2 cm}<{\centering}|m{2.2 cm}<{\centering}|m{1.2 cm}<{\centering}|c|m{3 cm}<{\centering}}
		\hline
		\textbf{} &  \textbf{DL-based comm.}& \textbf{DL-based sensing} &\textbf{Sensing-aided Comm.} & \textbf{Modular design}& \textbf{Transferability} & \textbf{Channel} \\
		\hline
		Proposed bedrock models&\CheckmarkBold&\CheckmarkBold&\CheckmarkBold&\CheckmarkBold&\CheckmarkBold&AWGN, doubly dispersive channel\\
		\hline
		Constellation mapping and AE design\cite{constellationdesign2}&\CheckmarkBold& & & & &AWGN, actual over-the-air channel\\
		\hline
		Pilot design and channel estimation\cite{pilotdesign}&\CheckmarkBold& & & & &Time dispersive channel\\
		\hline
		Constellation and filter design\cite{waveform2}&\CheckmarkBold& & & & &AWGN, 3GPP\\
		\hline
		Complex convolution for comm.\cite{CP_zhao}&\CheckmarkBold& & &\CheckmarkBold& &AWGN, Rayleigh\\
		\hline
		MUSIC enhancement\cite{Deep-MUSIC1}& &\CheckmarkBold& &\CheckmarkBold& &AWGN\\
		\hline
		Two network for comm. and sensing with flexible
		guard interval\cite{waveform_ISAC}&\CheckmarkBold&\CheckmarkBold&\CheckmarkBold& & &Doubly dispersive channel\\
		\hline
		Iterative optimization for sensing and comm.\cite{ISAC_NET}&\CheckmarkBold&\CheckmarkBold&\CheckmarkBold& & &Doubly dispersive channel\\
		\hline
		\end{tabular} \label{table:relatedwork}
	\end{table*}

	Despite these advancements, several challenges persist. First, many DL-based techniques for wireless communication and sensing tend to focus primarily on accuracy, often at the expense of generalization capabilities \cite{challenges}. Second, the sources of performance improvements in DL-based systems remain poorly understood, which hinders the interpretability of the models. Moreover, the E2E approach complicates the integration of DL-based models into existing communication and sensing systems.

	To address these challenges, we propose the concept of "bedrock models," which are grounded in the mathematical foundations of wireless communication and sensing. These models retain the modularity and generalization of both communication and sensing systems, thereby ensuring strong transferability. The key contributions of this work are as follows:

	\begin{itemize}
		
		\item We propose the bedrock models, which are based on mathematical formulations and a modular design approach. The bedrock model preserves generalization and transferability while offering flexibility to freely combine and replace traditional components, making it highly adaptable to existing communication and sensing systems.

		\item For communication systems, we design models for constellation mapping, DL-inverse discrete Fourier transform (IDFT)/discrete Fourier transform (DFT), and data extraction. Moreover, we analyze the feasibility of the proposed bedrock models, explained the design motivations, and provided performance guarantees. In the additive white Gaussian noise (AWGN) channel, the bit error rate (BER) performance of BPSK, QPSK, and 16-QAM modulation schemes is improved by approximately $2$ dB at low signal-to-noise ratio (SNR) and by $0.8$ dB at high SNR.
		
		\item Thanks to the modular design, we demonstrate the transferability of the DL-IDFT/DFT model. The results indicate that the DL-IDFT/DFT model trained with 16-QAM can be shared across BPSK and QPSK modulation tasks with negligible performance loss. This suggests that, for different modulation schemes, E2E transceiver training is not required. Instead, the receivers data extraction model can be trained with pilot signals, enabling more efficient system deployment.
		
		\item For sensing systems, we design three bedrock models for tasks as ECM estimation, noise subspace generation, and peak detection. Combined with traditional IDFT and DFT, we obtain interpretable results and enhance the performance of traditional MUSIC algorithm. In addition, we revisit the design motivations and performance guarantees based on the sensing framework. And we also identify the performance improvements introduced by each model in Section \ref{sec:performance4sensing}.
		
		\item We expand the sensing scheme and replace traditional IDFT/DFT models with the DL-IDFT/DFT models, achieving significant performance gains. At high SNR, delay and Doppler estimation errors can be reduced by an order of magnitude compared to traditional MUSIC algorithms. Moreover, the construction of the noise subspace remains unaffected by variations in SNR and is binarized. This means that after training, the model can be further simplified, reducing the complexity of transceiver design.
		
		\item We utilize sensing results to enhance communication performance in doubly dispersive channels. By designing a pre-equalization model within the bedrock framework, we integrate it at the transmitter to adjust the signal waveform based on delay and Doppler parameters. The results show that the communication system trained in an AWGN channel can be directly applied to a doubly dispersive channel without retraining, achieving BER performance comparable to the bedrock model in the AWGN channel.
	\end{itemize}

	The rest of this paper is organized as follows. Section \ref{sec:II} describes signal and channel models for both communications and sensing. Section \ref{sec:comms} presents the proposed bedrock models for communications. Section \ref{sec:sensing} delineates the proposed bedrock models for sensing. The application of bedrock models for sensing-aided communications and simulation results are elaborated in Section \ref{sec:sensing_aidedComm} and Section \ref{sec:simu_results}. Finally, the paper is concluded in Section \ref{sec:conclusion}.

	\emph{Notation:} The notation $[\cdot]_i$ and $[\cdot]_{i,j}$ refer to the $i$-th element of a vector and the $(i,j)$-th element of a matrix, respectively. The imaginary unit is denoted by $j=\sqrt{-1}$. The Euclidean norm and absolute value are denoted by $||\cdot||$ and $|\cdot|$, respectively. The operators $(\cdot)^T$ and $(\cdot)^H$ represent the transpose and conjugate transpose, respectively. The operator $\Re(\cdot)$ extracts the real component of a complex number, while $\Im(\cdot)$ extracts the imaginary component. Finally, the notation $\mathcal{CN}$ represents the circularly symmetric complex Gaussian distribution.

	\section{System Model}\label{sec:II}
	In this paper, we consider an OFDM-based system comprising a single-antenna transmitter that transmits OFDM signals for both communication and sensing purposes. The communication receiver is also equipped with a single antenna. We begin by introducing the OFDM signal and channel models that serve as the foundation for this work. Subsequently, we present the proposed bedrock models for OFDM-based communication and sensing in Sections \ref{sec:comms} and \ref{sec:sensing}, respectively. Finally, we demonstrate the effectiveness of the proposed bedrock models and extend them to a sensing-aided communication system in Section \ref{sec:sensing_aidedComm}.

	\subsection{Signal model}\label{sec:IIa}

	We consider an OFDM system, where each frame consists of $M$ consecutive OFDM symbols with a total of $N$ subcarriers within a coherent processing interval. To mitigate interference and maintain signal integrity, $N_g$ nullified guard subcarriers are positioned at the edges, while $N_d$ direct current (DC) subcarriers are placed at the center of each OFDM symbol. Additionally, $N_p$ pilot subcarriers, arranged in a comb pattern, are allocated for channel estimation. As a result, the number of subcarriers available for data transmission is $N-N_g-N_d-N_p$.

	Each subcarrier in an OFDM symbol is referred to as a resource element (RE), denoted as $S_{n,m}$, $n \in N$, $m \in M$. These symbols can belong to any constellation format, such as quadrature amplitude modulation (QAM). The subcarrier spacing, denoted as $\Delta f$, is defined as the inverse of the symbol duration $T$, i.e., $\Delta f = 1/T$. The length of the CP is denoted by $T_{cp}$, resulting in a total symbol duration of $T_o = T + T_{cp}$.
	
	Thus, the time-domain transmit signal $x(t)$ from the transmitter, including the CP, for both sensing and communication can be expressed as
	\begin{align}
		s(t)=\sum_{m=1}^{M} \sum_{n=1}^{N} S_{n, m} \operatorname{rect}\left(t-m T_o\right) e^{j 2 \pi n \Delta f\left(t-T_{cp}-m T_o\right)},
	\end{align}
	where $\operatorname{rect}(\cdot)$ denotes the rectangular shaping pulse with a duration $T$, used to modulate constellation symbols at the transmitter side, which can be express as
	\begin{align}
		\operatorname{rect}(t)=
		\begin{cases}
			1 & \text{if}\quad |t| \leq T / 2, \\
			0 & \text{otherwise}.
		\end{cases}
	\end{align}

	The CP, typically inserted to mitigate inter-symbol interference (ISI), is removed at the receiver before signal processing.

	\subsection{Channel model}\label{sec:IIb}
	We consider a time-varying doubly dispersive channel. For the sensing system, the channel impulse response of an $L$-path channel can be expressed as
	\begin{align}
		h^s(t, \tau) = \sum_{l=1}^L \beta^s_l  \delta\left(\tau-\tau^s_l\right) e^{j 2 \pi \nu^s_l t},
	\end{align}
	where $\beta^s_l$, $\tau^s_l$ and $\nu^s_l$ denotes the complex channel gain, round-trip delay and Doppler shift of $l$-th path, respectively. $f_c$ is the carrier frequency, and $c_0$ denotes the speed of the light. Thus, we have
	\begin{align}
		&\nu^s_l=\frac{2v^s_lf_c}{c_0},\nonumber\\
		&\tau^s_l = \frac{2r^s_l}{c_0},
	\end{align}
	where $v_l$ and $r_l$ denotes the speed and range of the $l$-th path. For the communication system, we have
	\begin{align}
		h^c(t, \tau) = \sum_{p=1}^P \beta^c_p  \delta\left(\tau-\tau^c_p\right) e^{j2\pi \nu^c_p t},
	\end{align}
	where $\beta^c_p$, $\tau^c_p$ and $\nu^c_p$ denotes the complex channel gain, one-way delay and Doppler shift of $p$-th path, and $\nu^c_l=v^c_lf_c/c_0$, $\tau^c_l=r^c_l/c_0$.
	
	By ignoring the noise for the sake of simplicity, the echo signal received by the transmitter can be express as
	\begin{align}
		Y^\S(t) = \int h^s(t, \tau) s(t-\tau) d \tau=\sum_{l=1}^{L}\beta^s_l s(t-\tau_l)e^{j2\pi \nu_l t}.
	\end{align}

	By sampling with period $T/N$ and CP removing, we obtain
	\begin{align}
		&Y_{n, m}^\S  \nonumber\\
		&=\left.Y^\S(t)\right|_{t=m T_o+T_{cp}+n T / N} \nonumber\\
		& =\sum_{l=1}^{L} \beta^s_l e^{j 2 \pi m T_o \nu_l} \sum_{n^{\prime}=1}^{N} S_{n, m^{\prime}} e^{j 2 \pi \frac{n}{N}\left(\frac{\nu_l}{\Delta f}+n^{\prime}\right)} e^{-j 2 \pi n^{\prime} \Delta f \tau_l}.
	\end{align}
	Typically, CP duration satisfies $T_{cp}\geq \tau_l^s, \forall l\in L$, ensuring that ISI is completely eliminated. Thus, after applying the DFT, we have\cite{OFDM_DDchannel}
	\begin{align}
		Y_{n, m} & =\frac{1}{N} \sum_{i=1}^{N} Y_{i, m}^\S e^{-j 2 \pi \frac{n i}{N}} \nonumber\\
		& \overset{(a)}{=}\frac{1}{N} \sum_{l=1}^{L} \beta^s_l e^{j 2 \pi \nu_l m T_o} \sum_{n^{\prime}=0}^{N-1} S_{n^{\prime}, m} e^{-j 2 \pi n^{\prime} \Delta f \tau_l} \nonumber\\
		& \qquad\qquad\qquad\qquad\qquad\quad\times \sum_{i=1}^{N} e^{j 2 \pi \frac{i}{N} \frac{\nu_l}{\Delta f}} e^{j 2 \pi \frac{i\left(n^{\prime}-n\right)}{N}} \nonumber\\
		& \overset{(b)}{\approx} \frac{1}{N} \sum_{l=1}^{L} \beta^s_l e^{j 2 \pi m T_o \nu_l} \sum_{m^{\prime}=0}^{M-1} S_{n^{\prime}, m} e^{-j 2 \pi n^{\prime} \Delta f \tau_l} \nonumber\\
		& \qquad\qquad\qquad\qquad\qquad\qquad\qquad\times\sum_{i=1}^{N} e^{j 2 \pi \frac{i\left(n^{\prime}-n\right)}{N}} \nonumber\\
		& \overset{(c)}{=} \sum_{l=1}^{L} \beta^s_l e^{j 2 \pi m T_o \nu_l} e^{-j 2 \pi n \Delta f \tau_l} S_{n, m}.\label{eqn:approximation_effect}
	\end{align}
	where $(b)$ follows from the assumption $\nu_l \ll \Delta f$, and $(c)$ exploits the orthogonality condition. Note that $(a)$ represents the actual channel, whereas $(b)$ and $(c)$ are approximations. In most of the literature and studies, the approximation in $(b)$ is commonly  adopted, leading to $(c)$, where delay and Doppler effects are treated separately\cite{DD_sep, DD_sep2}. In Section \ref{sec:sensing_aidedComm}, we further analyze the impact of this approximation and propose an adaptive improvement based on the bedrock models.

	\section{Proposed Bedrock Model for Communications}\label{sec:comms}
	\subsection{Bedrock Models Design}
	In this section, we first introduce the design of proposed bedrock models for OFDM communication systems in an AWGN channel for the sake of clarity. Specifically, we systematically design the constellation mapping, DFT/IDFT matrix and data demodulation models for communication. Subsequently, in Section \ref{sec:sensing_aidedComm}, we extend these models to a doubly dispersive channel by incorporating sensing results and channel pre-equalization.

\begin{figure}
		\centering
		\includegraphics[width=0.99\linewidth]{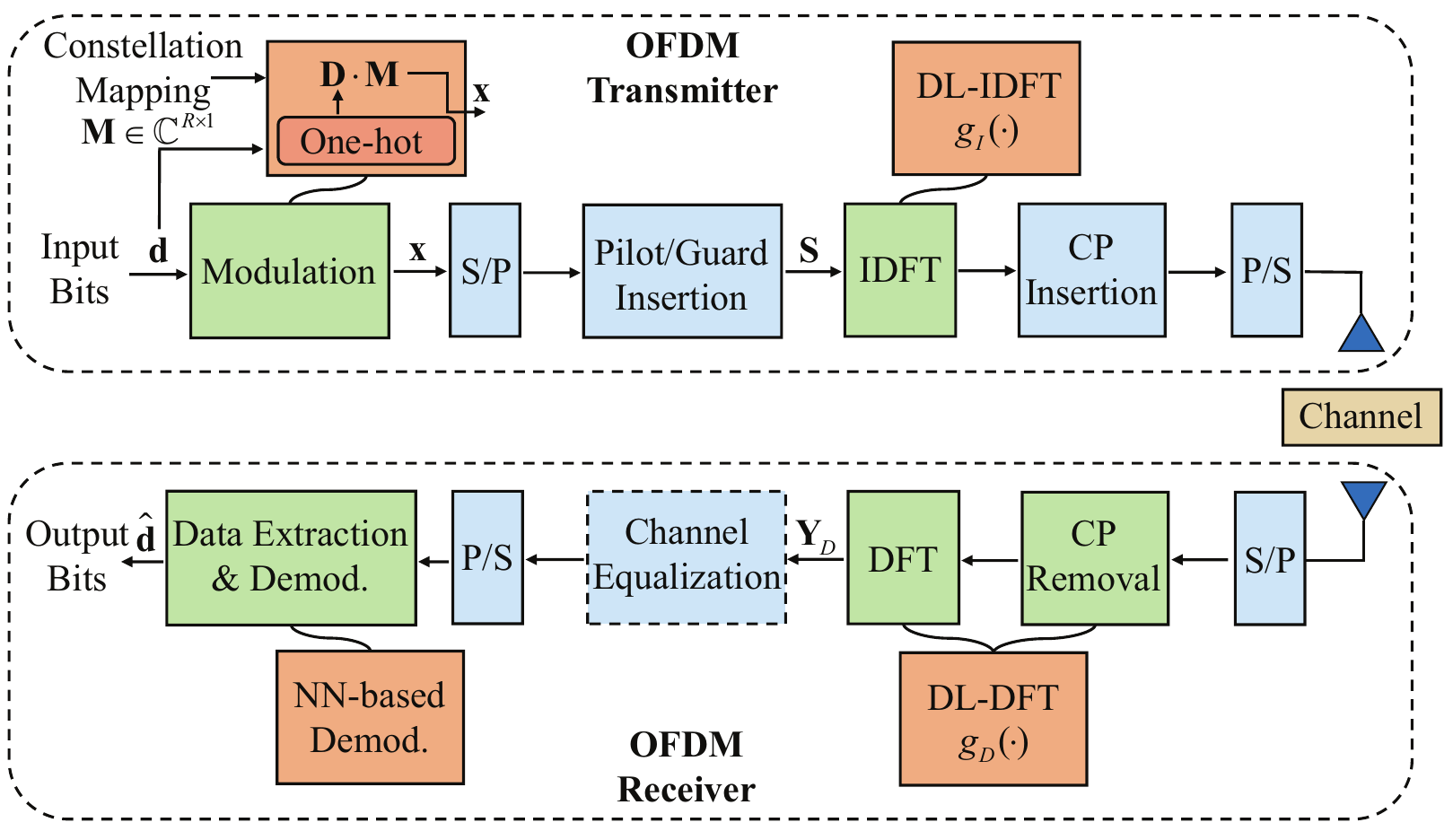}
		\caption{Diagram of OFDM physical layer and proposed bedrock models. We design the modulation (constellation mapping), DL-based IDFT/DFT and demodulation/data extraction models for communication enhancement.}
		\label{fig:commpipeline}
	\end{figure}

	Previous works have shown the potential for deep learning in communications\cite{OshePhysical, constellationNorm, coding1,DL_MIMO_OFDM_Comm}. As a bedrock design for communications, we rethink the system architectures by replacing the traditional models as shown in Fig. \ref{fig:commpipeline}. Note that these traditional models can be partially or completely replaced by our designed bedrock models, thereby increasing flexibility and achieving better performance while ensuring the architecture and transferability of traditional communication links.

	As the core component of communication systems, the modulation/constellation mapping model is considered first. Essentially, constellation mapping is the process of converting a sequence of digital bits into complex modulation symbols in digital communication. Thus, our goal is to design a trainable complex sequence which is more suitable for transmit with less BER via the channels, i.e., $\tilde{\mathbf{M}}=\left(\tilde{\mathbf{M}}_{real}+j\cdot\tilde{\mathbf{M}}_{imag}\right)\in\mathbb{C}^{R\times 1}$, where $R=2^r$ and $r$ is the modulation order, $\tilde{\mathbf{M}}_{real}\in\mathbb{R}^{R\times 1}$ and $\tilde{\mathbf{M}}_{imag}\in\mathbb{R}^{R\times 1}$ denote the real component and imaginary component of $\tilde{\mathbf{M}}$.

	The constellation mapping $\mathbf{M}$, which can be normalized as\cite{constellationNorm}
	\begin{align}
		\mathbf{M} = \frac{\tilde{\mathbf{M}}-\frac{1} {2^r}\sum_{\omega\in \tilde{\mathbf{M}}}\omega}{\sqrt{\frac{1}{2^r}\sum_{\omega\in \tilde{\mathbf{M}}}\left|\omega\right|^2-\left|\frac{1}{2^r}\sum_{\omega\in \tilde{\mathbf{M}}}\omega\right|^2}},
	\end{align}
	Thus, the trainable mapping points from the input symbols with one-hot representation $\mathbf{D}$ can be express as
	\begin{align}
		\mathbf{x} = \mathbf{D}\cdot \mathbf{M},
	\end{align}
	where $\mathbf{x}$ is the modulated complex baseband symbols.
	\begin{Rem}
		Note that many works have studied the design of constellation mapping, including the influence of geometric and probabilistic shaping \cite{ConsShaping}, coding \cite{coding1} and waveform \cite{constellationNorm, constellationdesign2}. However, we return to the essence by equating it to the design of complex sequence $\tilde{\mathbf{M}}$ and $\mathbf{M}$. This approach simplifies the model complexity and enables effective initialization, thereby accelerating the convergence speed. For instance, we can use traditional QAM-based constellation mapping as the initialization, incorporating characteristics such as Gray coding that are difficult to represent and train, thus reducing the difficulty of constellation training.
	\end{Rem}

\begin{figure*}[!t]
		\centering
		\subfloat[]{
			\includegraphics[width=0.19\linewidth]{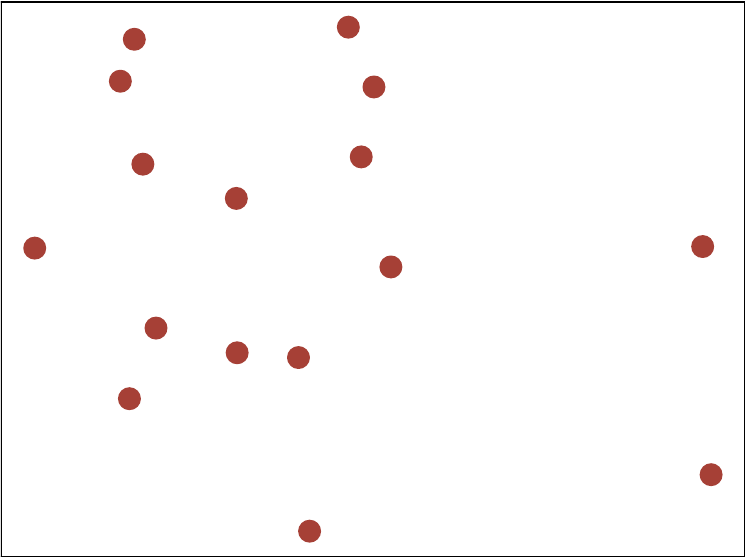}
		}
		\hspace{-2mm}
		\subfloat[]{
			\includegraphics[width=0.19\linewidth]{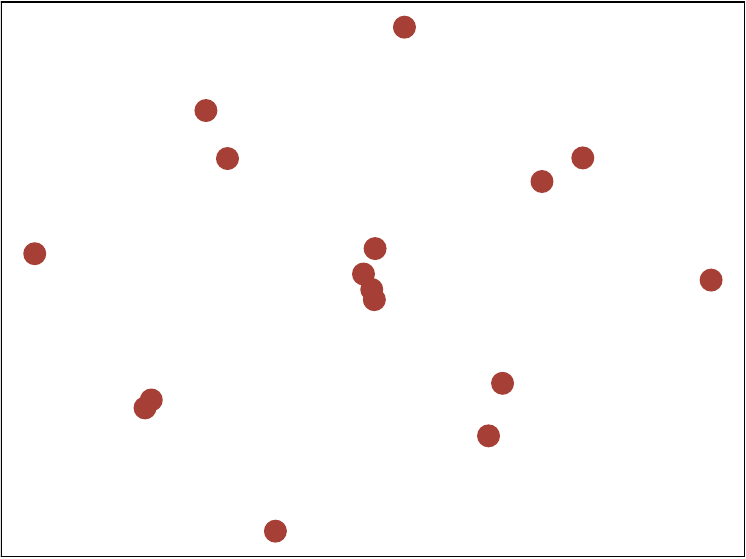}
		}
		\hspace{-2mm}
		\subfloat[]{
			\includegraphics[width=0.19\linewidth]{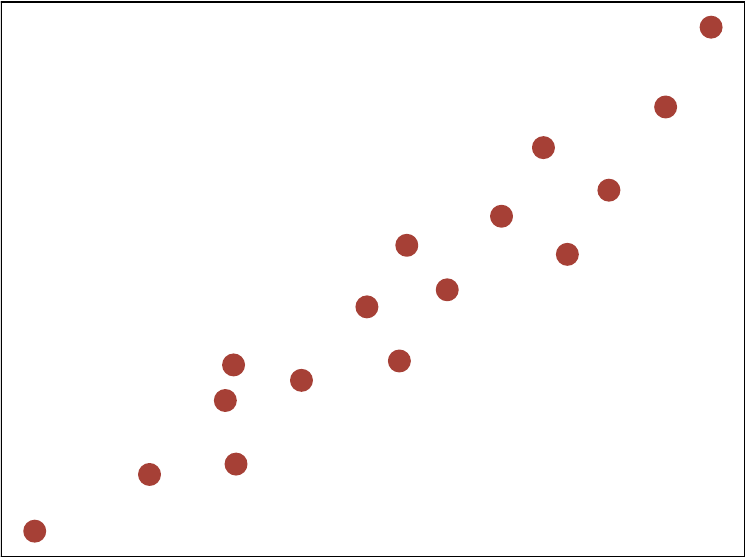}
		}
		\hspace{-2mm}
		\subfloat[]{
			\includegraphics[width=0.19\linewidth]{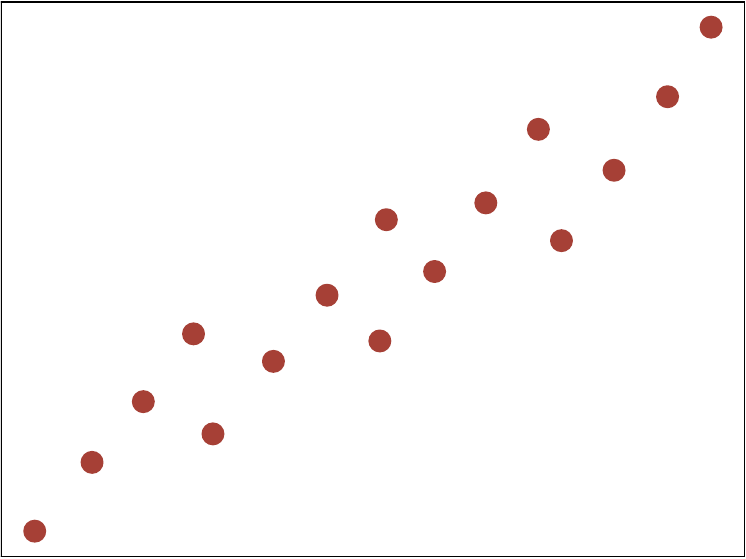}
		}
		\hspace{-2mm}
		\subfloat[]{
			\includegraphics[width=0.19\linewidth]{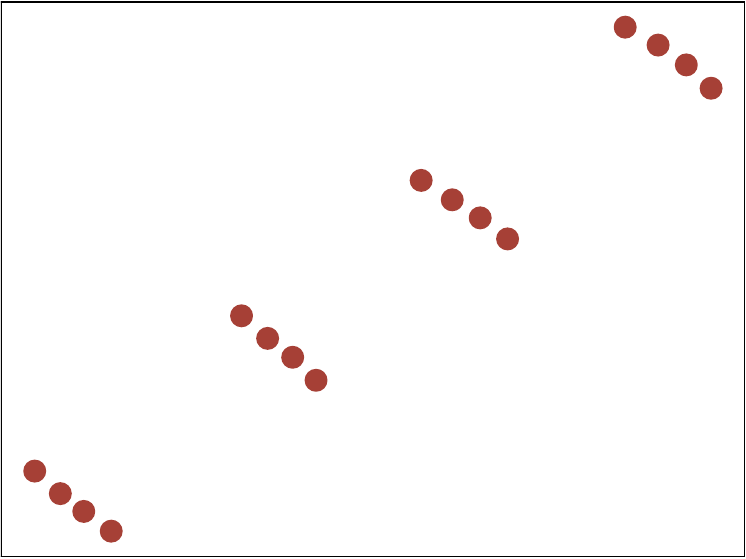}
		}
		\vspace{-2mm}
		\newline
		\centering
		\subfloat[]{
			\includegraphics[width=0.19\linewidth]{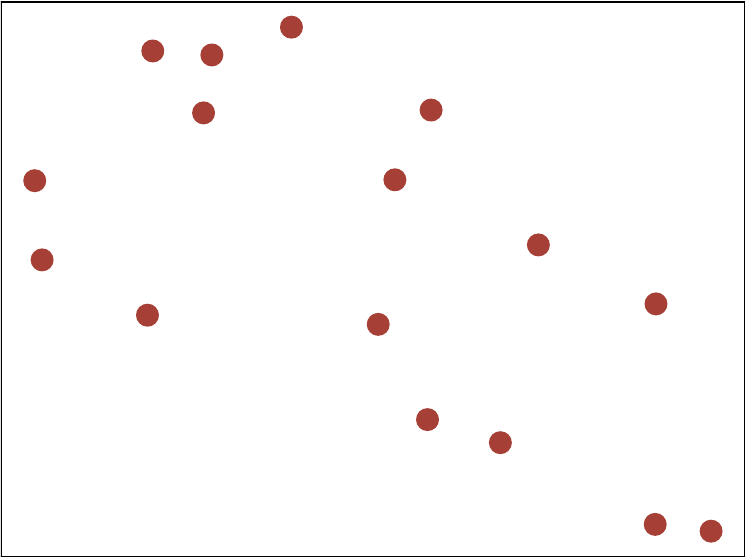}
		}
		\hspace{-2mm}
		\subfloat[]{
			\includegraphics[width=0.19\linewidth]{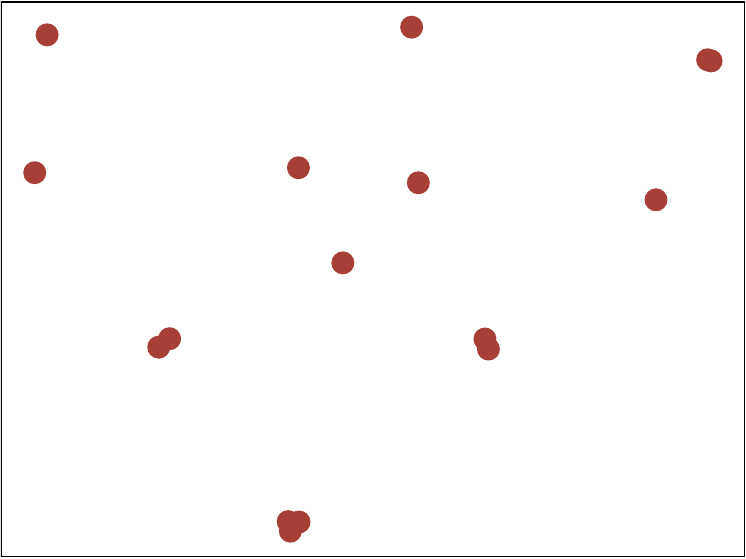}
		}
		\hspace{-2mm}
		\subfloat[]{
			\includegraphics[width=0.19\linewidth]{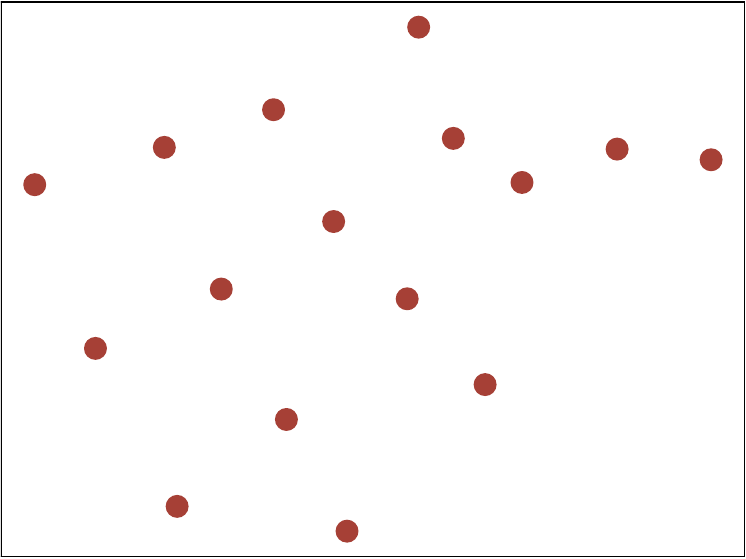}
		}
		\hspace{-2mm}
		\subfloat[]{
			\includegraphics[width=0.19\linewidth]{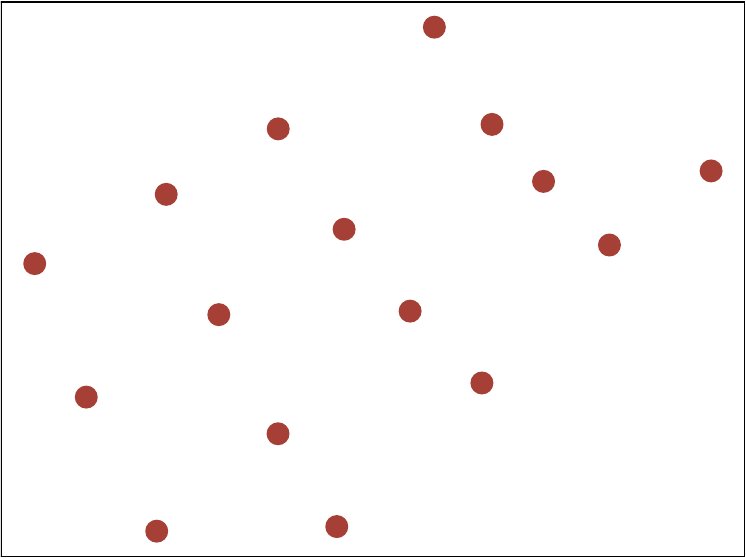}
		}
		\hspace{-2mm}
		\subfloat[]{
			\includegraphics[width=0.19\linewidth]{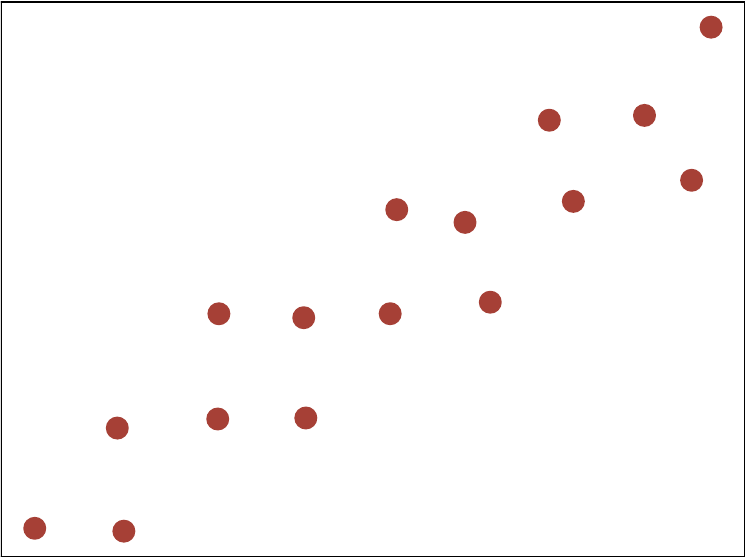}
		}
		\vspace{-2mm}
		\newline
		\centering
		\subfloat[]{
			\includegraphics[width=0.19\linewidth]{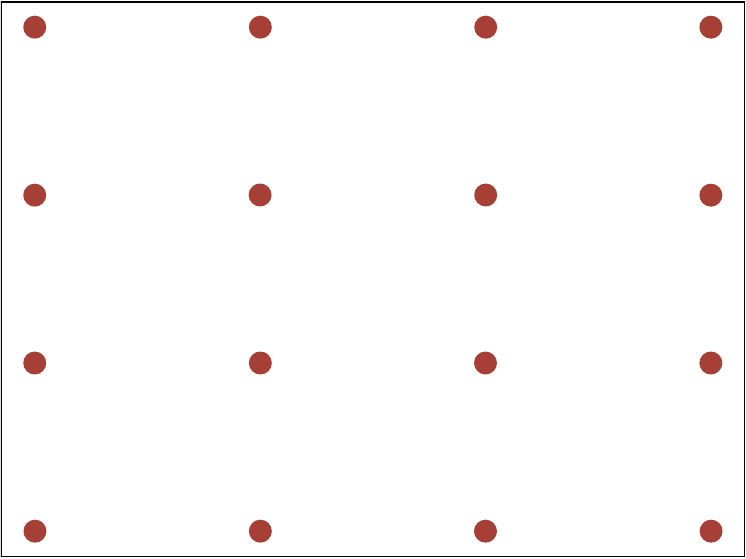}
		}
		\hspace{-2mm}
		\subfloat[]{
			\includegraphics[width=0.19\linewidth]{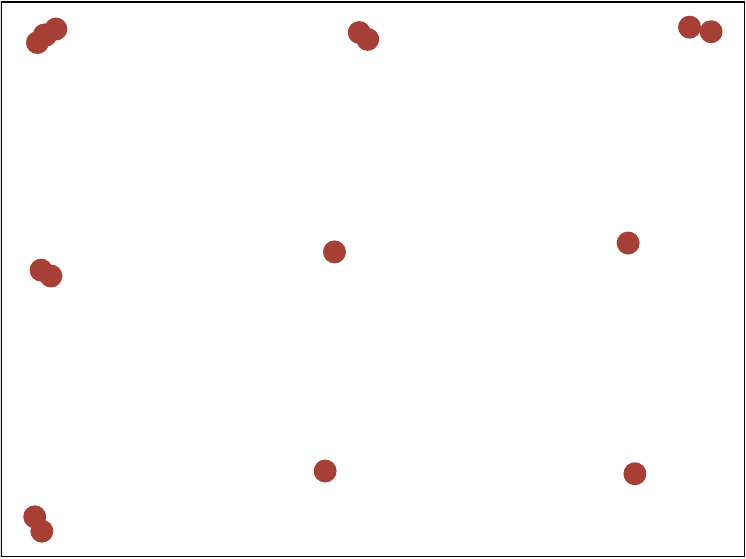}
		}
		\hspace{-2mm}
		\subfloat[]{
			\includegraphics[width=0.19\linewidth]{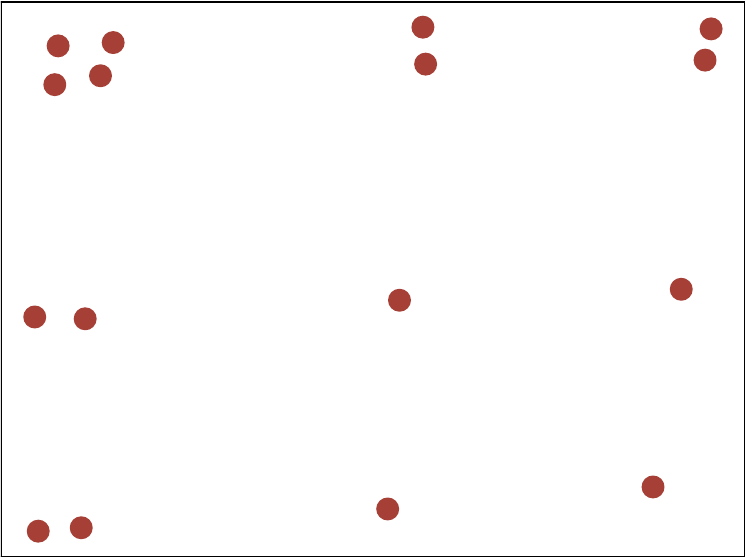}
		}
		\hspace{-2mm}
		\subfloat[]{
			\includegraphics[width=0.19\linewidth]{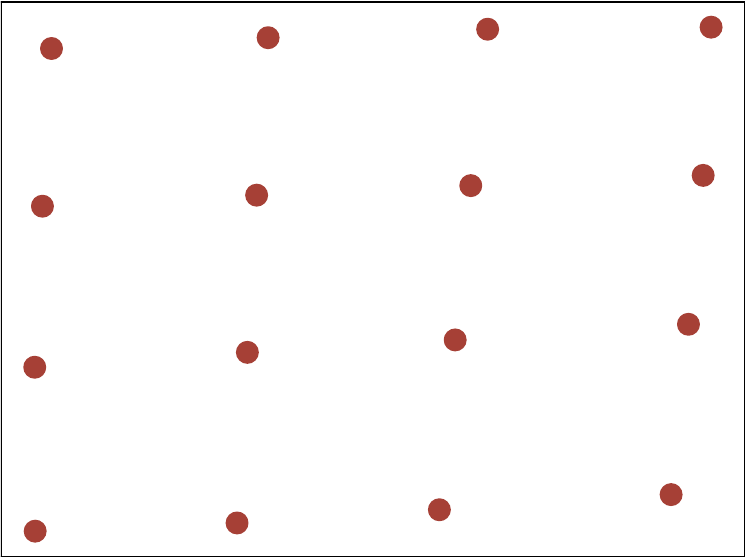}
		}
		\hspace{-2mm}
		\subfloat[]{
			\includegraphics[width=0.19\linewidth]{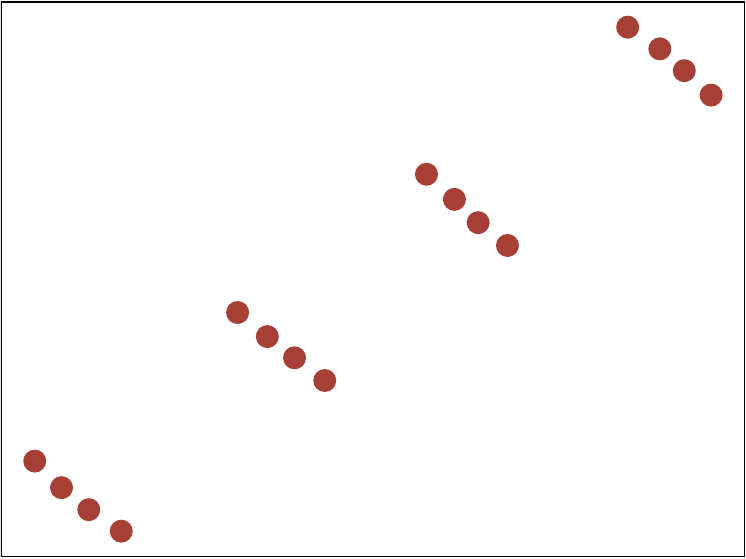}
		}
		\caption{$\text{(a)-(e)}$ shows the constellation training using two-layer NNs. while $\text{(f)-(j)}$ and $\text{(k)-(o)}$ illustrate the constellation training with $R\times 1$ complex sequences initialized using random and 16-QAM schemes, respectively.}
		\label{fig:constellation_training}
	\end{figure*}

	Fig. \ref{fig:constellation_training} illustrates the constellation training processes for NNs, random, and 16-QAM initialization schemes under a fixed SNR of 10 dB. Observed that all constellation points initially cluster to form a low-order modulation similar to QAM, before gradually spreading out. This behavior suggests that, during constellation training, the model seeks to group symbols with fewer adjacent bit errors together, such as symbols differing by only one bit, while increasing the distance between symbols that do not belong to the same cluster. This trend is particularly pronounced when training under very low SNR conditions. As training progresses, the constellation is further refined and separated. Furthermore, the random initialization scheme tends to converge to a local optimal solution, whereas the two-layer NN and 16-QAM initialization schemes are capable of reaching the global optimal solution\footnote{In fact, although both schemes are capable of achieving global optimality, the 16-QAM initialization scheme reduces the number of training iterations by 50\% compared to the NN scheme.}. However, it is important to note that training at only 10 dB resulted in overfitting of the constellation mapping, causing some constellation points to become too closely spaced.

	After constellation mapping and serial-to-parallel conversion, the pilot and guard symbols are inserted into the REs, as discussed in Section \ref{sec:IIa}. Next, we proceed to design the IDFT/DFT matrix. Traditional IDFT/DFT calculations can be expressed as
	\begin{align}
		\mathbf{Y} = \mathbf{F}_{DFT}^H\mathbf{Y}^\S,\nonumber\\
		\mathbf{Y}^\S = \mathbf{F}_{DFT}\mathbf{Y},
	\end{align}
	where $\left[\mathbf{Y}\right]_{n,m}=Y_{n,m}$, $\left[\mathbf{Y}^\S\right]_{n,m}=Y^\S_{n,m}$. And $\mathbf{F}_{DFT}\in\mathbb{C}^{N\times 1}$ denotes the DFT matrix. Thus, the complex DFT can be decomposed into two real operations as
	\begin{align}
		\Re\left\{\mathbf{Y}^\S\right\}=\Re\left\{\mathbf{F}_{DFT}\right\}\Re\left\{\mathbf{Y}\right\}-\Im\left\{\mathbf{F}_{DFT}\right\}\Im\left\{\mathbf{Y}\right\},\nonumber\\
		\Im\left\{\mathbf{Y}^\S\right\}=\Re\left\{\mathbf{F}_{DFT}\right\}\Im\left\{\mathbf{Y}\right\}+\Im\left\{\mathbf{F}_{DFT}\right\}\Re\left\{\mathbf{Y}\right\}.\label{eqn:DFT_like}
	\end{align}

\begin{figure}
		\centering
		\subfloat[Proposed DL-DFT model.]{
			\includegraphics[width=0.7\linewidth]{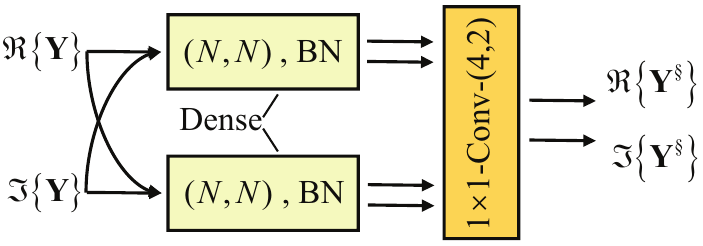}\label{fig:DFT-like}
		}
		\newline
		\centering
		\subfloat[Proposed data extraction/demodulation model.]{
			\includegraphics[width=0.99\linewidth]{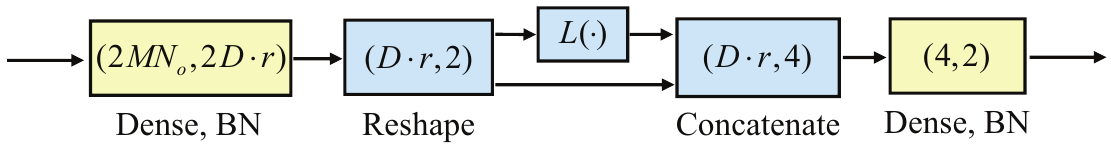}\label{fig:dataextra}
		}
		\caption{The flow graph of proposed models for communications, where $M$, and $N$ represent the frame size, and number of subcarriers, respectively. The label $(a, b)$ indicates the corresponding dense layer has an input dimension of $a$ and output dimension of $b$. Batch normalization (BN) is employed to stabilize the training process. $D=\left(N-N_g-N_d-N_p\right)\times M$ and $r$ denote the number of REs in a frame and the order of modulation, respectively, and $L(\cdot)$ denotes the LeakyReLU activation function.}
	\end{figure}

	Eq. \eqref{eqn:DFT_like} motivates the design of a one-layer NN and a $1\times 1$ convolutional neural network (CNN), referred to as the DL-DFT and denotes as $g_D(\cdot)$, to simulate the complex operations as shown in Fig. \ref{fig:DFT-like}. The real and imaginary components of $\mathbf{Y}$ are passed through a dense layer, after which the four outputs are concatenated. The final real and imaginary components are then obtained through a $1\times 1$ CNN.
	
	\begin{Rem}\label{remark:2}
		Specifically, we use the dense layer for matrix multiplication, such as $\mathbf{a}\Re\left\{\mathbf{Y}\right\}+\mathbf{b}$, to simulate the process $\Re\left\{\mathbf{F}_{DFT}\right\}\Re\left\{\mathbf{Y}\right\}$. We then apply a $1\times 1$ CNN, with an input channel size of 4 and an output channel size of 2, to perform addition and subtraction as described in Eq. \eqref{eqn:DFT_like}. It is important to note that the dense layer, such as $\mathbf{a}\Re\left\{\mathbf{Y}\right\}+\mathbf{b}, \mathbf{b}\neq \mathbf{0}$, represents a non-linear transformation, even without a non-linear activation function, as it does not satisfy the additivity property of linear transformations. This characteristic enhances the expressive power of the DL-DFT model. However, it also introduces certain challenges, which we will discuss further in Section \ref{sec:sensing_aidedComm}. Moreover, in contrast to traditional communication schemes, the DL-DFT/IDFT-based system allows for the retention of the CP at the receiver, enabling the extraction of useful information for enhanced performance. In this case, the dimension $N$ of the DL-DFT at the receiver is extended to $N_o=N+N_{cp}$, where $N_{cp}$ represents the length of the CP.
	\end{Rem}

	As the final component of the communication system, we design a NN-based framework for data extraction and demodulation, as illustrated in Fig. \ref{fig:dataextra}. The input data, consisting of $MN_o=M(N+N_{cp})$ information REs, is initially converted into a real tensor by concatenating the real and imaginary components. This tensor is then passed through a linear layer with BN for data extraction. Afterward, it is reshaped into tensor dimensions of $[B, D\cdot r, 2]$, followed by the application of a LeakyReLU activation function. The activated data is subsequently concatenated into a tensor of shape $[B, D\cdot r, 4]$, which is then fed into the final linear layer for demodulation. A sigmoid activation function is applied along the last dimension to generate logical values, effectively transforming the task into a binary classification problem. The binary cross-entropy (BCE) loss, used to quantify the discrepancy between the original bits $\mathbf{d}$ and their predicted counterparts $\hat{\mathbf{d}}$, is expressed as

	\begin{align}
		\mathcal{L}_{comm}=\sum^B\sum_{i=1}^{D\cdot r}\left(d_i \log \hat{d}_i+\left(1-d_i\right) \log \left(1-\hat{d}_i\right)\right),
	\end{align}
	where $d_i \in \mathbf{d}$ and $\hat{d}_i \in \hat{\mathbf{d}}$ represent the input bits and the corresponding prediction bits, respectively.

	In this section we introduce three key components of the proposed bedrock model for communications: constellation mapping, DL-DFT/IDFT, and data extraction/demodulation design. Thanks to its modular design, the proposed bedrock model can be seamlessly integrated with or substituted for traditional communication models. Furthermore, the model exhibits several notable features, including improved performance by CP remain, and transferability. For further details, please refer to Section \ref{sec:performance4comm}.
	
	\subsection{Feasibility Analysis}\label{sec:FA_comm}

	It is essential to emphasize the underlying motivation for the design of the bedrock models. The objective is to develop a model that not only possesses a modular architecture but also ensures transferability and superior performance. In this context, the design of the bedrock models for communication tasks presented in this paper is rooted in mathematical principles. This approach offers the advantage of framing traditional schemes as special cases within our bedrock models. Specifically, the constellation mapping model, which is trainable and operates within a geometric space of $R$ complex points, inherently includes traditional schemes as feasible solutions. This is further demonstrated in Fig. \ref{fig:constellation_training}.

	For the DL-IDFT/DFT model, the special case occurs when
	\begin{align}
		\mathbf{Y}^\S_1 = \mathbf{a}_{1}\Re\{\mathbf{Y}\}+\mathbf{b}_1=\Re\left\{\mathbf{F}_{DFT}\right\}\Re\left\{\mathbf{Y}\right\},\nonumber\\
		\mathbf{Y}^\S_2 = \mathbf{a}_{2}\Re\{\mathbf{Y}\}+\mathbf{b}_2=\Im\left\{\mathbf{F}_{DFT}\right\}\Re\left\{\mathbf{Y}\right\},\nonumber\\
		\mathbf{Y}^\S_3 = \mathbf{a}_{1}\Im\{\mathbf{Y}\}+\mathbf{b}_1=\Re\left\{\mathbf{F}_{DFT}\right\}\Im\left\{\mathbf{Y}\right\},\nonumber\\
		\mathbf{Y}^\S_4 = \mathbf{a}_{2}\Im\{\mathbf{Y}\}+\mathbf{b}_2=\Im\left\{\mathbf{F}_{DFT}\right\}\Im\left\{\mathbf{Y}\right\},
	\end{align}
	where $\mathbf{a}_1\mathbf{x}+\mathbf{b}_1$ and $\mathbf{a}_2\mathbf{x}+\mathbf{b}_2$ denote the two Dense layers demonstrated in Fig. \ref{fig:DFT-like}, respectively. Thus, we have 
	\begin{align}
		\mathbf{a}_1=\Re\left\{\mathbf{F}_{DFT}\right\}, \mathbf{b}_1=\mathbf{0},\nonumber\\
		\mathbf{a}_2=\Im\left\{\mathbf{F}_{DFT}\right\}, \mathbf{b}_2=\mathbf{0}.
	\end{align}
	Moreover, as the four input channels of $1\times 1$ CNN, $\mathbf{Y}^\S_1$, $\mathbf{Y}^\S_2$, $\mathbf{Y}^\S_3$ and $\mathbf{Y}^\S_4$ are processed with the weights of $[1, 0, 0, -1]$ and $[0, 1, 1, 0]$ for DL-DFT, and $[1, 0, 0, 1]$ and $[0, -1, 1, 0]$ for DL-IDFT, respectively, and obtain a two-channel output. As a result, the DF-DFT/IDFT operation becomes fully equivalent to traditional DFT/IDFT according to Eq. \eqref{eqn:DFT_like}, with the functions $g_D(\cdot)$ and $g_I(\cdot)$ exhibit linear transformations. Finally, the data extraction model serves as the final classification model, leveraging the powerful fitting capabilities of NNs to adaptively process features from either traditional models or those based on the bedrock framework, ultimately delivering the final decision.

	This design ensures that by replacing traditional models with any of the proposed bedrock models, the performance of the communication system will be at least as good as, or better than, that of the traditional scheme. Since each bedrock model incorporates the corresponding traditional scheme as one of its feasible solutions, thereby guaranteeing a performance lower bound.

	\section{Proposed Bedrock Model for Sensing}\label{sec:sensing}
	\subsection{Bedrock Model Design}
	In this section, we describe the problem of sensing tasks in doubly dispersive channel, including problem formulation and sensing parameters estimation, and followed with the proposed bedrock model for better performance.
	
	For the sake of clarity, we set $L=1$ and the received signals after DFT can be simplified as
	\begin{align}
		Y_{n,m} = \beta^s_0 S_{n, m}e^{j 2 \pi m T_o \nu_0} e^{-j 2 \pi n \Delta f \tau_0}+z_{n,m},
	\end{align}
	where $z_{n,m}$ denotes the additive white Gaussian noise. We have $z_{n,m}\sim \mathcal{CN}\left(0, \sigma_0^2\right)$ and $\sigma_0^2$ denotes the noise power. Thus, By taking the Hadamard product of the received signal and conjugate of the transmitted signal, the phase impact caused by symbol $S_{n,m}$ can be eliminated, resulting in
	\begin{align}
		&R_{n,m}=Y_{n,m}\odot S^\dagger_{n,m} \nonumber\\
		&= \beta^s_0 \left|S_{n, m}\right|^2 e^{j 2 \pi m T_o \nu_0} e^{-j 2 \pi n \Delta f \tau_0}+\left|S_{n, m}\right|z_{n,m},\label{eqn:symbolelimi}
	\end{align}
	where $\odot$ denotes Hadamard product. $S_{n,m}^\dagger z_{n,m}=\left|S_{n,m}\right|e^{-j2\pi\phi_{n,m}}z_{n,m}$, and $\phi_{n,m}$ denotes the phase introduced by $S_{n,m}^\dagger$. Therefore, the power of $z_{n,m}$ becomes $\sigma_0^2e^{-j2\pi\phi_{n,m}}e^{j2\pi\phi_{n,m}}=\sigma_0^2$, indicating that the phase $\phi_{n,m}$ does not affect the noise power\footnote{Note that $z_{n,m}\neq z_{n,m}e^{-j2\pi\phi_{n,m}}$, but its statistical characteristics remains unchanged. Thus, we continue to use $z_{n,m}$ to represent the corresponding noise.}. As a result, the SNR remains unchanged when multiplying both the signal and noise by $\left|S_{n,m}\right|$ simultaneously.

	Thus, the covariance matrix $\mathbf{E}_{\tau_0}=\mathbf{R}\mathbf{R}^H\in\mathbb{C}^{N\times N}$ and $\mathbf{E}_{\nu_0}=\mathbf{R}^H\mathbf{R}\in\mathbb{C}^{M\times M}$, where $[\mathbf{R}]_{n,m}=R_{n,m}$, can be expressed as
	\begin{align}
		&\mathbf{E}_{\tau_0} = \nonumber\\
		&\begin{bmatrix}
			A^\tau_{1,1} & \dots & A^\tau_{1,N} e^{j 2 \pi (N-1) \Delta f \tau_0} \\
			A^\tau_{2,1} e^{-j 2 \pi \Delta f \tau_0} & \dots & A^\tau_{2,N} e^{j 2 \pi (N-2) \Delta f \tau_0} \\
			\vdots & \ddots & \vdots \\
			A^\tau_{N,1} e^{-j 2 \pi (N-1) \Delta f \tau_0} & \dots & A^\tau_{N,N}\label{eqn:cov4music}
		\end{bmatrix},\\
		&\mathbf{E}_{\nu_0} = \nonumber\\
		&\begin{bmatrix}
			A^\nu_{1,1} & \dots & A^\nu_{1,M} e^{j 2 \pi (M-1) T_o \nu_0} \\
			A^\nu_{2,1} e^{-j 2 \pi T_o \nu_0} & \dots & A^\nu_{2,M} e^{j 2 \pi (M-2) T_o \nu_0} \\
			\vdots & \ddots & \vdots \\
			A^\nu_{M,1} e^{-j 2 \pi (M-1) T_o \nu_0} & \dots & A^\nu_{M,M}
		\end{bmatrix},
	\end{align}
	by ignoring the noise for the sake of simplicity, where $A^\tau_{i,j}=\left(\beta^s_0\right)^2\sum_{m=1}^M\left|S_{i, m}\right|^2\left|S_{j, m}\right|^2,\forall i,j\in N$, and note that $A^\tau_{i,j}=A^\tau_{j,i},\forall i,j\in N$. Similarly, $A^\nu_{i,j} = \left(\beta^s_0\right)^2\sum_{n=1}^N\left|S_{n, i}\right|^2\left|S_{n, j}\right|^2,\forall i,j\in M$. Thus, the delay and Doppler estimation from $\mathbf{E}_{\tau_0}$ and $\mathbf{E}_{\nu_0}$ can be viewed as a classic frequency estimation problem, which can be effectively solved using super-resolution algorithms such as multiple signal classification (MUSIC) algorithm\cite{MUSIC}.

	Specifically, considering delay estimation as an example, parameter estimation using the MUSIC algorithm can be separated into three main steps \cite{Deep-MUSIC1, Deep-MUSIC2}. First, the ECM $\mathbf{E}_{\tau_0}$ is derived from Eq. \eqref{eqn:cov4music}. Then, the signal subspace $ \mathbf{U}_{\tau_0}^{s} = [\mathbf{U}_{\tau_0}]_{:, 1} \in \mathbb{C}^{N \times 1} $ and the noise subspace $ \mathbf{U}_{\tau_0}^{n} = [\mathbf{U}_{\tau_0}]_{:, 2:} \in \mathbb{C}^{N \times (N-1)} $ are extracted from the $\mathbf{E}_{\tau_0}$ via eigendecomposition, under the assumption that the signal and noise are uncorrelated. Here, $\mathbf{U}_{\tau_0}$ represents the matrix of eigenvectors corresponding to the eigenvalues, which are arranged in descending order. The eigenvectors associated with the dominant eigenvalues span the signal subspace, while those corresponding to the smaller eigenvalues (i.e., $N-1$ for a single delay estimation) form the noise subspace.

	Finally, the pseudo spectrum is expressed as:

	\begin{align}
		P\left(\tau\right) = \frac{1}{\boldsymbol{\alpha}^H\left(\tau\right)\mathbf{U}_{\tau_0}^n\left(\mathbf{U}_{\tau_0}^n\right)^H\boldsymbol{\alpha}\left(\tau\right)}, \label{eqn:pesudoSpec}
	\end{align}

	where $\boldsymbol{\alpha}(\tau)$ denotes the linear phase and is given by $\boldsymbol{\alpha}(\tau) = \left[ 1, e^{-j2\pi \Delta f \tau}, \dots, e^{-j2\pi (N-1) \Delta f \tau} \right]^T$. The parameter $\tau$ must traverse the scanning range, and due to the orthogonality between the noise subspace and the signal subspace, the function $P(\tau)$ exhibits a peak at $\tau \approx \tau_0$.

\begin{figure}
		\centering
		\includegraphics[width=0.99\linewidth]{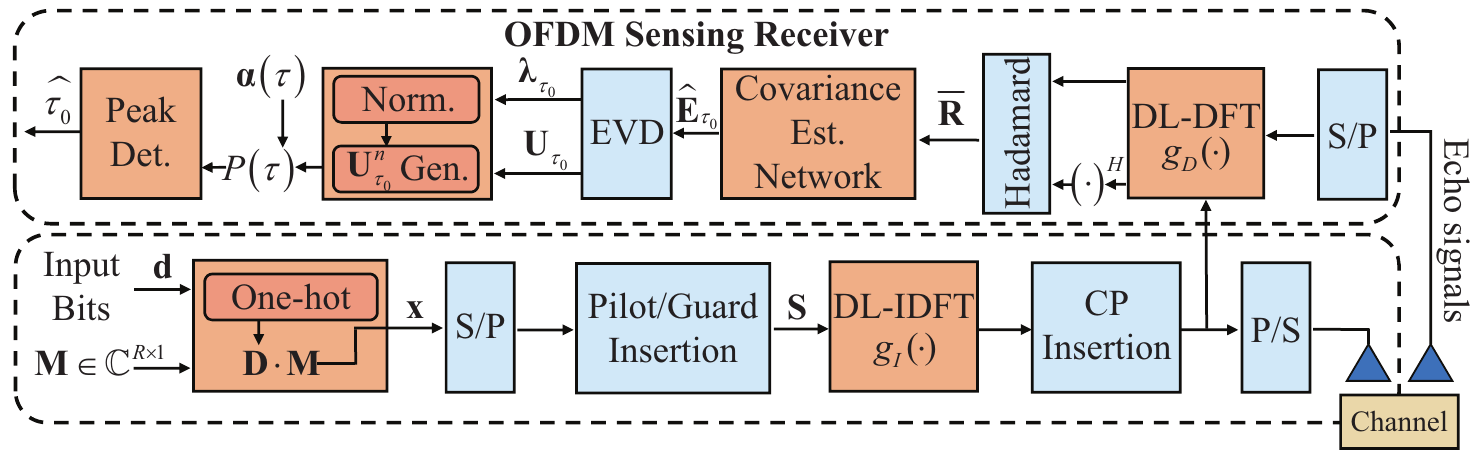}
		\caption{Diagram of OFDM sensing and proposed bedrock models. We design three models for sensing enhancement: the ECM estimation model, the noise subspace generation model, and the peak detection model.}
		\label{fig:sensingpipeline}
	\end{figure}

	To further enhance the performance of parameter estimation while leveraging the same models from the communication system, we propose bedrock sensing models, as illustrated in Fig. \ref{fig:sensingpipeline}. As discussed in Remark \ref{remark:2}, the non-linear and non-orthogonal transformation DL-DFT/IDFT, denoted as $g_D(\cdot)$ and $g_I(\cdot)$, are adopted for performance enhancement. Thus, we modify the calculation in Eq. \eqref{eqn:symbolelimi} to allow the transmitted data to undergo a non-linear transformation by $g_D(\cdot)$, followed by a Hadamard product with the received signal, thereby more effectively mitigating the influence of transmit symbols\footnote{More generally, the traditional scheme in Eq. \eqref{eqn:symbolelimi} can be considered as a special case of our modification. Specifically, after performing IDFT and DFT on $S_{n,m}$, the orthogonality of IDFT and DFT ensures that $S_{n,m}$ remains unchanged. However, we do not impose orthogonality on $g_I(\cdot)$ and $g_D(\cdot)$, allowing the process to remain intact.}. Once the value of $\bar{\mathbf{R}}$ is obtained from Eq. \eqref{eqn:symbolelimi}, a covariance estimation network is employed to ensure that $\boldsymbol{\alpha}(\tau)$ lies within the signal subspace of the estimated covariance matrix. Then, motivated by \cite{eigVectorSelection}, a partial noise subspace is utilized for parameter estimation using a selector model, which generates the subspace $\mathbf{U}_{\tau_0}^n$. Finally, leveraging the ability of NNs to provide continuous results, we replace the grid-on output of the traditional MUSIC algorithm with a three layer NNs for continuous output, enabling a grid-free estimation.

\begin{figure}
		\centering
		\subfloat[Proposed ECM estimation model.]{
			\includegraphics[width=0.25\linewidth]{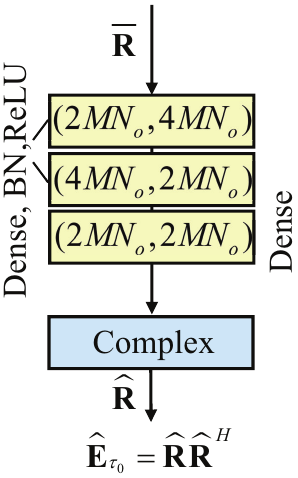}\label{fig:empnet}
		}
		\hspace{2mm}
		\subfloat[Proposed partial noise subspace selection model.]{
			\includegraphics[width=0.65\linewidth]{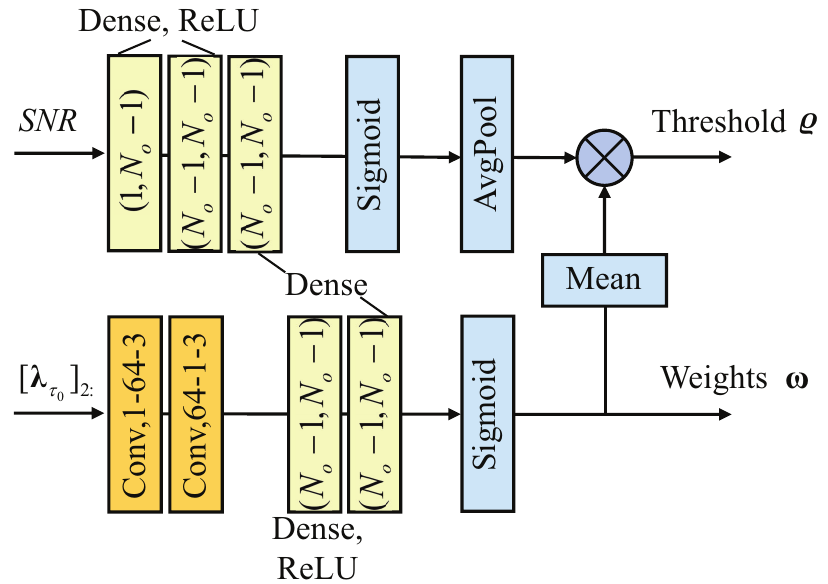}\label{fig:weightnet}
		}
		\caption{The flow graph of proposed models for sensing, where $N_o=N+N_{cp}$. The label $\textit{\text{Conv}-a-b-c}$ indicates that the corresponding convolutional layer has an input dimension of $a$, an output dimension of $b$, and a kernel size of $c$. The label $\textit{\text{Average}}$ denotes the average operation along the last dimension.}
	\end{figure}

	The covariance estimation network, which can obtain a better ECM and construct a more accuracy noise subspace, is shown in Fig. \ref{fig:empnet}. Specifically, the input complex data $\bar{\mathbf{R}}$ is reshaped into $[B, MN_o]$, and the real and imaginary components are concatenated to form a real tensor of shape $[B, 2MN]$. After passing through two dense layers, the input is separated into real and imaginary components and reshaped into $[B, N, M]$. The resulting output $\hat{\mathbf{R}}$ is then used to compute $\hat{\mathbf{E}_{\tau_0}}$ via $\hat{\mathbf{E}}_{\tau_0}=\hat{\mathbf{R}}\hat{\mathbf{R}}^H$.

	Motivated by \cite{eigVectorSelection}, we construct a partial noise subspace to achieve better parameter estimation performance. In fact, selecting eigenvectors for noise subspace is equivalent to multiplying the eigenvectors by binarized weights. Therefore, after performing eigenvalue decomposition (EVD), we first sort the eigenvalues in descending order and normalize them according to the dominant eigenvalue. Then, we select the eigenvalues, excluding the dominant eigenvalue, i.e., $[\boldsymbol{\lambda}_{\tau_0}]_{2, :}$, and SNR, as inputs, and determine their corresponding weights for the eigenvectors. To ensure the binarization of the weights, we design the binary weights as
	\begin{align}
		\boldsymbol{\omega}_b=\text{Sigmoid}\left(\eta\left(\boldsymbol{\omega}-\boldsymbol{\varrho} \right)\right),\label{eqn:weights}
	\end{align}
	where $\eta$ is hyperparameter that controls the rate at which the binarization occurs. The non-binary weights $\boldsymbol{\omega}$ and threshold $\boldsymbol{\varrho}$ can be obtain by the model shown in Fig. \ref{fig:weightnet}.

	And the selected $\mathbf{U}^n_{\tau_0}$ can be expressed as
	\begin{align}
		\mathbf{U}^n_{\tau_0} = \left\{\left[\mathbf{U}_{\tau_0}\right]_{:, i}\cdot[\boldsymbol{\omega}_b]_{i}|1\leq i\leq N_o-1\right\},
	\end{align}
	where $[\boldsymbol{\omega}_b]_{i}\in\left\{0, 1\right\}$ as $\eta$ is sufficiently large and the number of training iterations is sufficiently high.

	By constructing the pseudo-spectrum as shown in Eq. \eqref{eqn:pesudoSpec}, the final step involves continuous output generation on the grid through a peak detection model. Using a three-layer Dense network with Tanh activation functions, we obtain the normalized predicted results. The normalized prediction is then multiplied by the maximum estimated range to obtain the final grid-free output, i.e., $\hat{\tau_0}$.

	To evaluate the performance, the loss function is computed by comparing the parameters estimated from the sensing bedrock models with the true parameters, such as $\hat{\tau_0}$ and $\tau_0$. Therefore, the mean square error (MSE) is employed as the loss function, which can be expressed as

	\begin{align}
		\mathcal{L}_{sens}=\frac{1}{B}\sum^B\left|\hat{\tau_0}-\tau_0\right|^2+\frac{1}{B}\sum^B\left|\hat{\nu_0}-\nu_0\right|^2.
	\end{align}

	\subsection{Feasibility Analysis}
	
	For the sensing bedrock models, due to the nonlinear nature of $g_D(\cdot)$, the constructed vector $\boldsymbol{\alpha}(\tau_0)$ may not necessarily be orthogonal to the noise subspace $\mathbf{U}^n_{\tau_0}$. However, a special case of the DL-IDFT/DFT, as discussed in Section \ref{sec:FA_comm}, demonstrates that when $\mathbf{b}_i = 0,\forall i\in \{1,2\}$, the transformation $\mathbf{a}_i \mathbf{x}, \forall i \in \{1, 2\}$, undergoes a linear transformation. This property ensures that $\boldsymbol{\alpha}(\tau)$ remains within the signal subspace, thereby preserving the effectiveness of the algorithm. Similarly, the ECM estimation model, as an additional module, not only enhances sensing models based on the bedrock framework but also improves traditional scheme by smoothing and strengthening the signals, thereby achieving more accurate ECM estimation.

	Furthermore, the noise subspace generation model, due to the binarization design in Eq. \eqref{eqn:weights}, forces the weighting coefficients of the eigenvectors to either 0 or 1, effectively selecting or discarding the eigenvectors. Let $\mathbf{U}^n_{\tau_0}$ denotes the partial noise subspace constructed by selecting $K$ eigenvectors from $\mathbf{U}^n_{\tau_0, total}$. Based on the properties of the signal and noise subspaces, we have
	\begin{align}
		\boldsymbol{\alpha}(\tau_0) \perp \mathbf{U}^n_{\tau_0, total},
	\end{align}
	which also means $\boldsymbol{\alpha}(\tau_0)\cdot [\mathbf{U}^n_{\tau_0, total}]_{:, i}=0, \forall i\in N$. Thus, we can further obtain 
	\begin{align}
		\boldsymbol{\alpha}(\tau_0) \perp \mathbf{U}^n_{\tau_0},
	\end{align}
	since $[\mathbf{U}^n_{\tau_0}]_{:, k} \in \mathbf{U}^n_{\tau_0, total}$ and $\boldsymbol{\alpha}(\tau_0)\cdot [\mathbf{U}^n_{\tau_0}]_{:, k}=0,\forall k\in K$. Note that this conclusion can be easily extended to arbitrary non-binarized weights, demonstrating that the proposed noise subspace model retains orthogonal with constructed signal vector $\boldsymbol{\alpha}(\tau_0)$. By combining the final grid-free NNs output, which can adaptively receive pseudo-spectra from either traditional models or those based on the bedrock framework, providing continuous estimation results.
	
	The sensing design based on bedrock models is analogous to the communication design, retaining the mathematical structure of traditional models. This ensures that replacing traditional algorithms with bedrock models does not degrade performance. This further emphasizes the underlying motivation behind our proposed bedrock model design.

	\section{Applications for Integrated Sensing and Communications}\label{sec:sensing_aidedComm}
	In this section, we introduce a pre-equalization schemes for sensing-aided communication system in doubly dispersive channel, by leveraging the bedrock design of communication and sensing models. Typically, the channel equalization process is applied at the receiver end as shown in Fig. \ref{fig:commpipeline}. However, this scheme requires the receiver to capture channel state information (CSI), which increases its computational complexity at the receiver end. Another feasible solution is to perform pre-equalization at the transmitter. Through the combined effects of the transmitter and the channel, the receiver can ideally recover to a time-invariant channel under optimal conditions\cite{DDA_1, DDA_2}. In traditional communication schemes, the approximate expression of the received signal after DFT can be found in Eq. \eqref{eqn:approximation_effect}. Based on Eq. \eqref{eqn:approximation_effect}, we introduce pre-equalization after the CP insertion, aiming to mitigate the effects of delay and Doppler shift as much as possible at the receiver.
	
\begin{figure}
		\centering
		\subfloat[$\nu_0=1.87$KHz]{
			\includegraphics[width=0.3\linewidth]{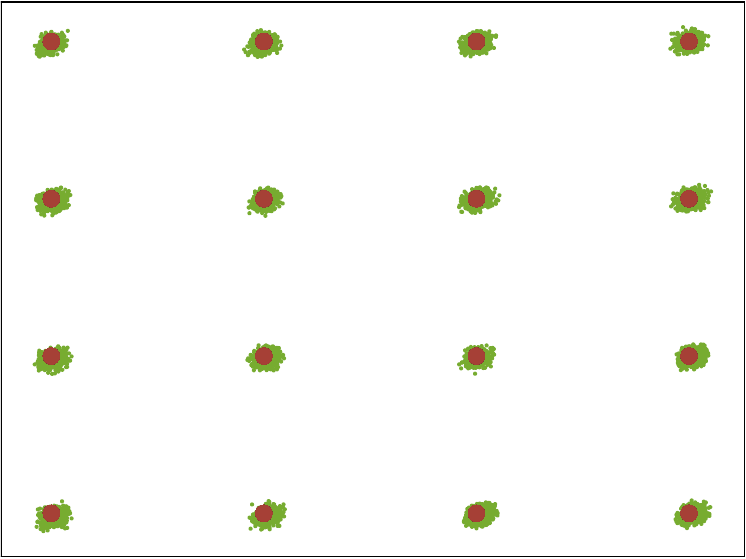}
		}
		\hspace{-2mm}
		\subfloat[$\nu_0=9.34$KHz]{
			\includegraphics[width=0.3\linewidth]{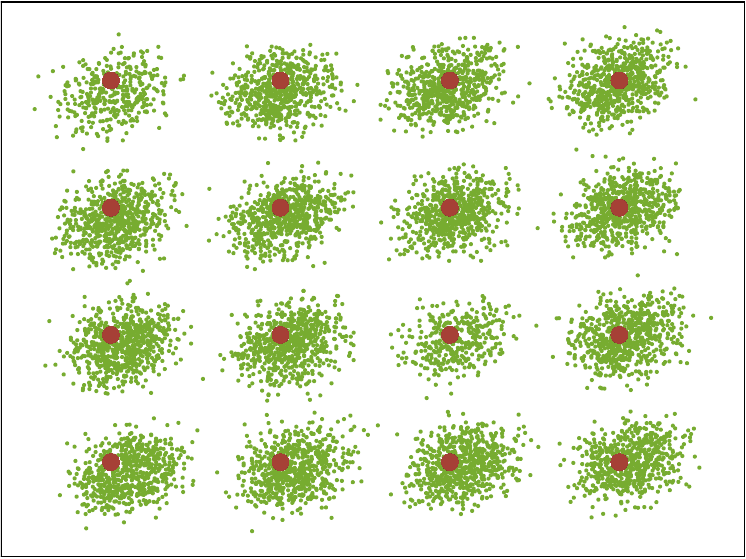}
		}
		\hspace{-2mm}
		\subfloat[$\nu_0=18.68$KHz]{
			\includegraphics[width=0.3\linewidth]{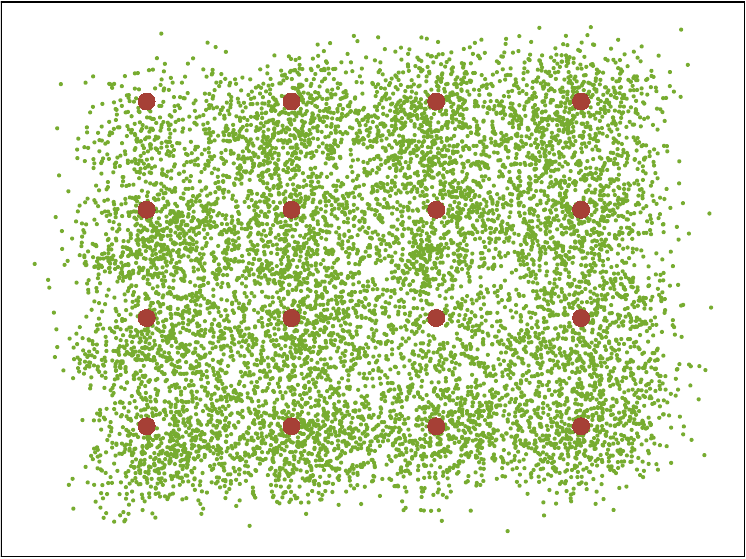}
		}
		\caption{Constellation mapping with ICI. The subcarrier spacing $\Delta f=120$KHz, $N=64$, $N_{cp}=16$, $M=8$ and carrier frequency $f_c=28$GHz. Thus, the Doppler $\nu_0/\Delta f=\{0.0156, 0.0778, 0.1557\}$.}
		\label{fig:const_velo}
	\end{figure}

	An intuitive approach is to multiply the transmitted signal $S_{n,m}$ by corresponding $e^{-j 2\pi m T_o\nu^c_p},\forall p\in P$ and $e^{j2\pi n \Delta f\tau^c_p},\forall l\in L$. Thus, the received signals at communication receiver with one-way delay and Doppler shift can then be rewritten as
	\begin{align}
		\hat{Y}_{n, m}&\approx \sum_{p=1}^{P} \beta^c_p e^{j 2 \pi m T_o \nu^c_p} e^{-j 2 \pi n \Delta f \tau^c_p} \nonumber\\
		&\qquad\qquad\qquad\times S_{n, m}e^{-j 2\pi m T_o\nu^c_p}e^{j2\pi n \Delta f\tau^c_p}\nonumber\\
		&= \sum_{p=1}^{P} \beta^c_pS_{n, m}.\label{eqn:equ_approx}
	\end{align}
	However, two main issues need to be further considered.
	\begin{enumerate*}
        \item As noted in Remark \ref{remark:2}, the proposed DL-IDFT/DFT does not satisfy orthogonality, which leads to the invalidation of $(c)$ in Eq. \eqref{eqn:approximation_effect}.
		\item Moreover, the approximation in $(b)$ of Eq. \eqref{eqn:approximation_effect} also needs to be taken into account.
    \end{enumerate*}
	Even under traditional IDFT and DFT, Doppler frequency shift induces inter-carrier interference (ICI), which can degrade communication performance. This point is further illustrated in Fig. \ref{fig:const_velo}.

	Fig. \ref{fig:const_velo} shows the demodulated constellation mapping at the receiver with noiseless channel, where $\tau_0=0$ and $\nu_0=\{1.87, 9.34, 18.68\}$KHz, corresponding to the radial velocity $v_0=\{10, 50, 100\}$ meters per second (m/s). Note that due to the neglect of ICI, the communication performance with pre-equalization using Eq. \eqref{eqn:equ_approx} will be severely affected, particularly when $\nu_0/\Delta f=0.1557$. This indicates that additional processing is required to ensure reliable communication performance.

\begin{figure}
		\centering
		\includegraphics[width=0.99\linewidth]{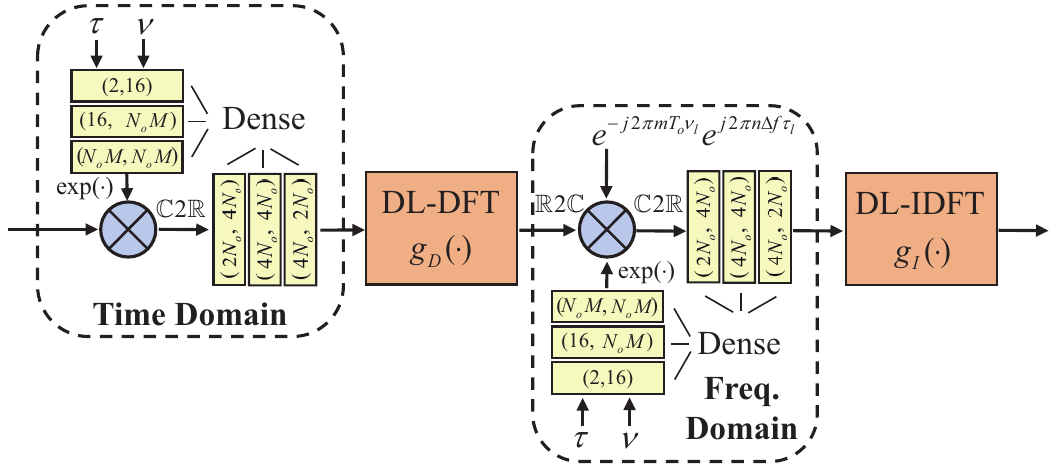}
		\caption{Diagram of proposed pre-equalization model for sensing-aided communications.}
		\label{fig:preequ}
	\end{figure}

	In response, we design a pre-equalization model, as depicted in Fig. \ref{fig:preequ}, which is positioned after the CP insertion block at the OFDM transmitter. Specifically, based on the sensing parameters, i.e., delay and Doppler, a dual-domain pre-equalization scheme is designed, operating in both the time and frequency domains. In the time domain, delay and Doppler are used to extract features through a three-layer dense model, which are then multiplied with the signal and passed through another three-layer dense network similar to ECM estimation model for equalization. After the time-domain equalization is performed, the signal is transformed to the frequency domain via DL-DFT\footnote{Although DL-IDFT/DFT is not a true time-frequency domain transformation, we maintain this terminology for simplicity}. Equalization is then performed in the frequency domain in a manner analogous to that in the time domain. It is important to note that the equalization method from Eq. \eqref{eqn:equ_approx} is retained in the frequency domain, specifically by multiplying by the exponential term $e^{-j 2\pi m T_o\nu_l},\forall l\in L$ and $e^{j2\pi n \Delta f\tau_l},\forall l\in L$, to achieve faster convergence. Finally, the signal is transformed back to the time domain through DL-DFT for transmission.

	\begin{table}
		\centering
		\caption{The Default System Parameters.}
		\begin{tabular}{c<{\centering}|c<{\centering}|m{3cm}<{\centering}}
		\hline
		\textbf{Parameters Description} &  \textbf{Notation} &\textbf{Default Values}\\
		\hline
		Number of subcarriers & $N$ & $64$\\
		CP length& $N_{cp}$ & $16$\\
		Pilot REs&$N_p$& $8$\\
		Guard REs& $N_{g}$ &$2\times 4$\\
		DC REs&$N_{dc}$ & $2$\\
		Data REs per frame&$D$ & $368$\\
		OFDM symbols per frame&$M$& $8$\\
		Modulation order &$R$& QAM-like with orders of 2, 4, 16\\
		Pilot type & -& Constant value $\sqrt{1/2}(1+j)$, with comb pattern\\
		Carrier frequency & $f_c$& $28$GHz\\
		Subcarrier spacing & $\Delta f$& $120$Khz\\
		Distance range & - & Uniform distribution from 0 to 100 meters \\
		Velocity range & - & Uniform distribution from 0 to 100 m/s \\
		MUSIC grid points &-& 100\\
		\hline
		Learning rate &-& 0.001\\
		Batchsize &-& 1000\\
		\hline
		\end{tabular} \label{table:systemParas}
	\end{table}
	\section{Simulation Results}\label{sec:simu_results}
	In this section, we first demonstrate the performance of bedrock models for communication, including performance enhancement and transferability. Then, sensing results are presented to compare the performance of traditional MUSIC algorithm and shown the power of designed bedrock models for sensing. Furthermore, the pre-equalization models are then applied for communications in doubly dispersive channel, without any retrain or finetune of the transmitter and receiver trained in AWGN channel. The default system parameters are detailed in Table \ref{table:systemParas}.

	\subsection{Performance for Communications}\label{sec:performance4comm}
\begin{figure}
		\centering
		\includegraphics[width=0.99\linewidth]{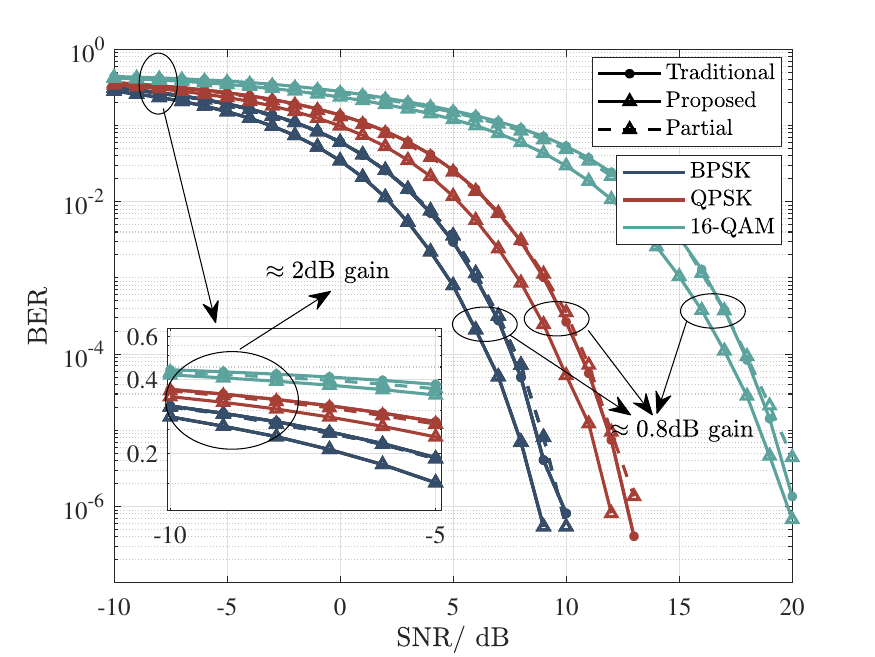}
		\caption{BER performance of traditional, bedrock and partial models.}
		\label{fig:exp1_berAWGN}
	\end{figure}

	In this section, we first consider the model introduced in Section \ref{sec:comms} with an AWGN channel. Both the transmitter and receiver, including constellation mapping, DL-IDFT/DFT, and data extraction, are strung together and trained as an E2E system. Fig. \ref{fig:exp1_berAWGN} shows the BER for $r = 1, 2, 4$-order QAM modulation using the proposed bedrock model, along with the corresponding traditional modulation, IDFT/DFT, and data extraction models.

	In this section, we first analyze the model introduced in Section \ref{sec:comms} under AWGN channel. The entire system, including the transmitter and receiver components, is integrated and trained as an E2E framework. Fig. \ref{fig:exp1_berAWGN} presents the BER performance for $r = 1, 2, 4$-order QAM modulation using the proposed bedrock models, alongside the corresponding conventional modulation, IDFT/DFT, and data extraction.

	\begin{figure}
		\centering
		\subfloat[BPSK-like]{
			\includegraphics[width=0.3\linewidth]{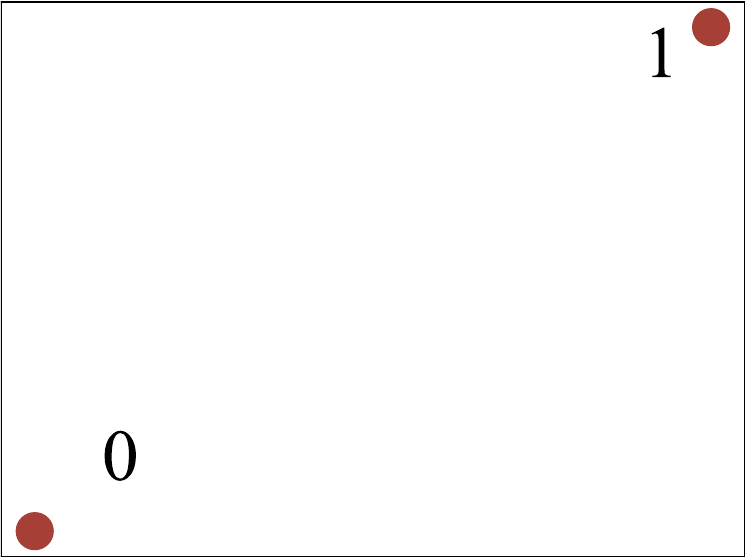}
		}
		\hspace{-2mm}
		\subfloat[QPSK-like]{
			\includegraphics[width=0.3\linewidth]{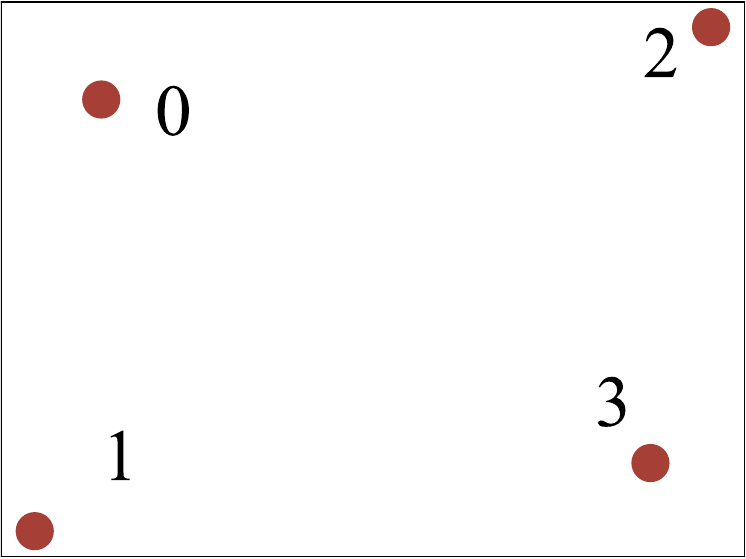}
		}
		\hspace{-2mm}
		\subfloat[16-QAM-like]{
			\includegraphics[width=0.3\linewidth]{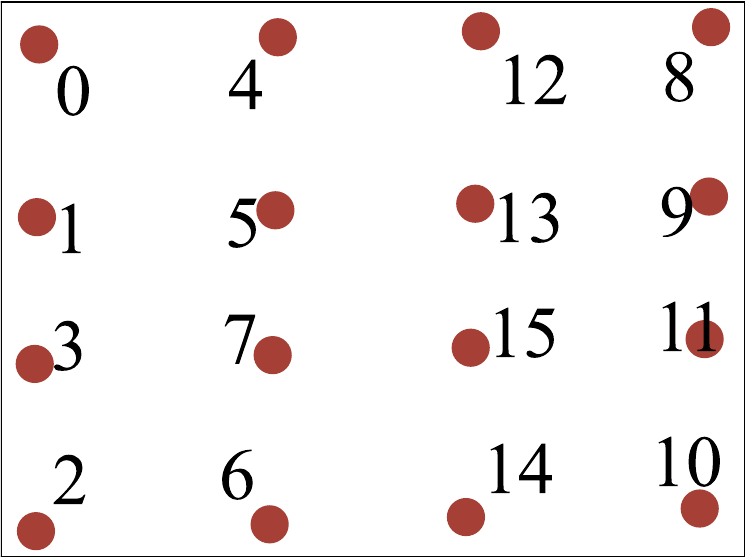}
		}
		\caption{Constellation points generated by the constellation mapping model for BPSK, QPSK and 16-QAM. There is only a 1-bit difference between adjacent symbols.}
		\label{fig:AWGN_constellation}
	\end{figure}

	Observed that the communication system based on the proposed bedrock model exhibits outstanding performance in the AWGN channel, showing a gain of approximately $2$ dB in BER at low SNR and a gain of about $0.8$ dB even at high SNR. Furthermore, due to the modular design, we replaced the constellation mapping and DL-IDFT models with traditional counterparts, retaining only the DL-DFT and data extraction models. The corresponding BER results are also presented in Fig. \ref{fig:exp1_berAWGN} for a more detailed analysis. This reveals that the performance improvement in AWGN primarily stems from the design of the constellation mapping and DL-IDFT models at the transmitter. Moreover, Fig. \ref{fig:AWGN_constellation} presents the constellation points generated by the constellation mapping model. Note that the resulting constellation naturally exhibits properties similar to Gray coding, wherein adjacent symbols differ by only 1 bit.

\begin{figure}
		\centering
		\subfloat[BER performance of model parameters sharing.]{\label{fig:exp2_para_sharing}
			\includegraphics[width=0.99\linewidth]{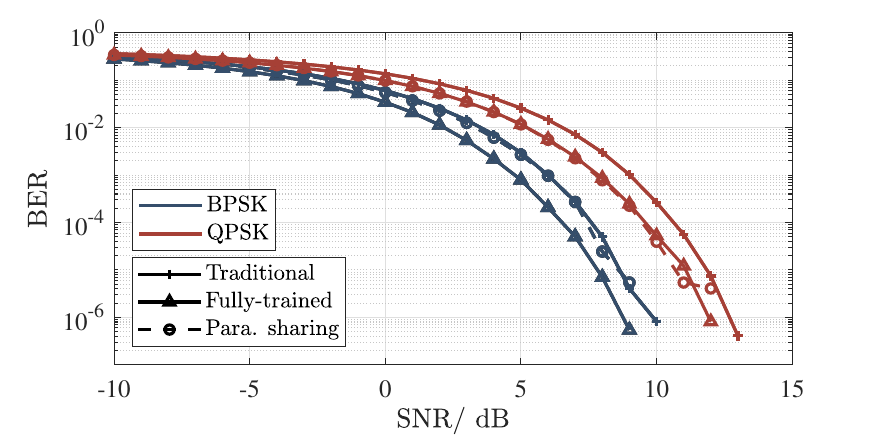}
		}
		\newline
		\subfloat[Loss values of model parameters sharing.]{\label{fig:exp2_loss}
			\includegraphics[width=0.99\linewidth]{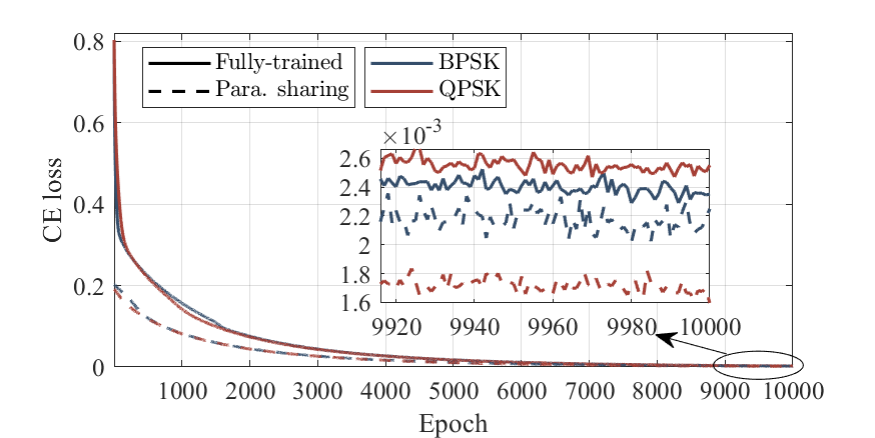}
		}
		\caption{BER performance and loss values of DL-IDFT/DFT model parameters sharing.}
		\label{fig:exp2accandloss}
	\end{figure}

	Another notable feature of the bedrock model in communication systems is its transferability. We share the DL-IDFT/DFT model from the 16-QAM models for training BPSK and QPSK, by freezing its parameters while only training the constellation mapping and data extraction models for BPSK and QPSK. The communication performance results following parameter sharing are shown in Fig. \ref{fig:exp2_para_sharing}. It can be observed that, with parameter sharing, QPSK achieves the same performance as a fully-trained bedrock model. However, a slight degradation in BPSK performance is noted at high SNR, with its performance falling between traditional QPSK and BPSK.

	Nonetheless, the benefits of model parameter sharing are evident. Fig. \ref{fig:exp2_loss} illustrates the variations in the loss function during both the parameter sharing and fully-trained models, clearly demonstrating that with parameter sharing and reuse, the loss function converges more rapidly. Moreover, this implies that when the modulation order is changed at the transmitter, such as from 16-QAM to QPSK, the receiver can achieve performance nearly equivalent to that of a fully-trained model by training only the data extraction model at the receiver using pilot signals. This capability is not typically available in other E2E communication models. And this compatibility and transferability offer substantial advantages for practical deployment and implementation.

	\subsection{Performance for Sensing}\label{sec:performance4sensing}

	To better illustrate the performance gain of the bedrock model for sensing tasks, we first retain the traditional IDFT/DFT components, combined with the proposed noise subspace generation and peak detection models, to achieve enhanced performance. Additionally, we consider scenarios both with and without the DC and guard subcarriers, allowing for a more effective demonstration of the performance\footnote{Notably, removing the DC and guard subcarriers leads to a modification of the vector $\boldsymbol{\alpha}(\tau)$ in Eq. \eqref{eqn:pesudoSpec} during delay parameter estimation, transforming it into a non-uniform phase vector corresponding to the data subcarriers.}.

\begin{figure}
		\centering
		\subfloat[Delay estimation training with $0\sim9$ dB.]{\label{fig:Delay_DCguard}
			\includegraphics[width=0.99\linewidth]{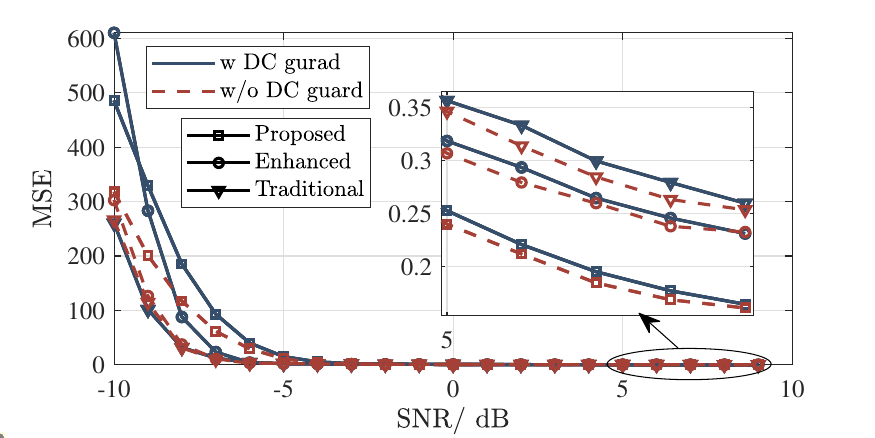}
		}
		\newline
		\subfloat[Doppler estimation training with $0\sim9$ dB.]{\label{fig:Doppler_DCguard}
			\includegraphics[width=0.99\linewidth]{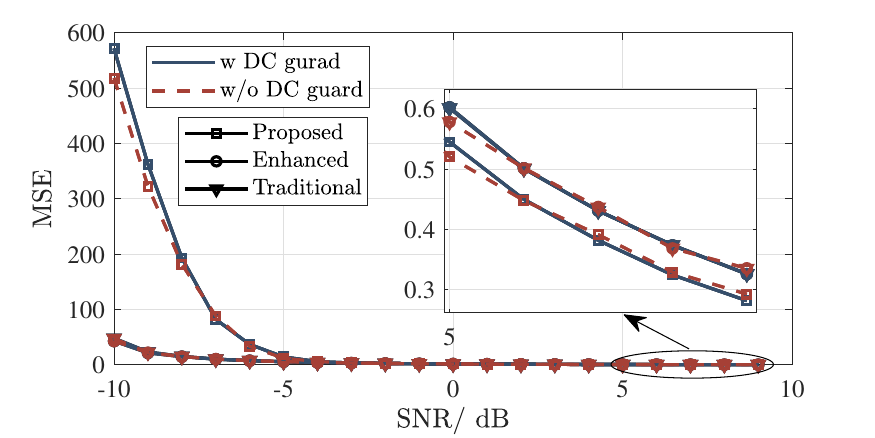}
		}
		\caption{Parameters estimation performance with proposed noise subspace generation and peak detection models. The results are obtained by training the model within the $0\sim9$ dB SNR range and $\eta = 10$.}
		\label{fig:MSEdelaydoppler}
	\end{figure}

	Fig. \ref{fig:Delay_DCguard} and Fig. \ref{fig:Doppler_DCguard} demonstrate the parameters estimation performance of both proposed bedrock model, enhanced and traditional models,  where $\eta = 10$. Note that within the training SNR range, i.e., $0\sim9$ dB, the proposed model achieved superior performance. Furthermore, the enhanced model outperforms the traditional model but falls slightly short compared to the proposed model, highlighting the effectiveness of the noise generation and peak detection models. Observed from Fig. \ref{fig:Delay_DCguard} that removing the DC and guard subcarriers can improve performance slightly in delay estimation. However, removing the DC and guard subcarriers appears to have no impact on Doppler estimation. This finding can reduce computational complexity and improve parameter estimation performance.
	
	However, the results outside the training SNR range, i.e., $-10\sim0$ dB, exhibit the opposite trend, with the traditional model gaining an advantage. To address this issue, we recommend training the proposed model separately for low SNR to achieve better performance\footnote{It is important to note that we do not suggest expanding the training SNR range to use a single model. This is because, at very low SNR, excessively large loss values may lead the model to neglect performance at high SNR, thereby degrading performance in the high SNR regime.}.

\begin{figure}
		\centering
		\subfloat[Delay]{
			\includegraphics[width=0.75\linewidth]{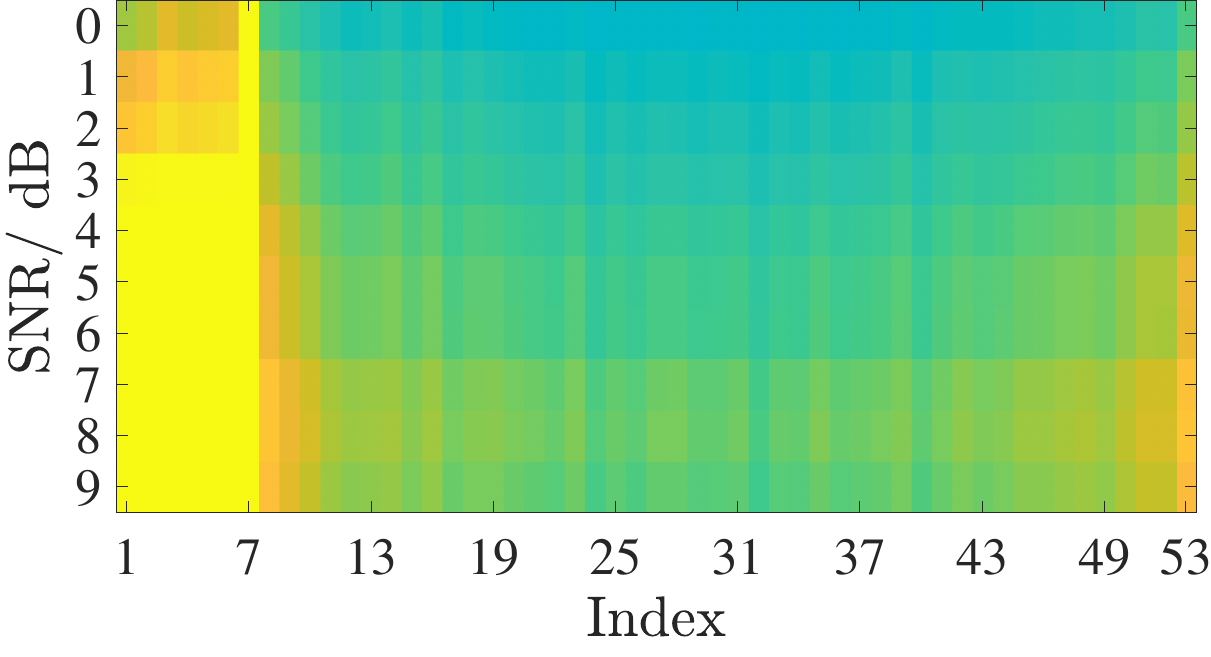}
		}
		\hspace{-2mm}
		\subfloat[Doppler]{\label{fig:wDCguardDoppler}
			\includegraphics[width=0.16\linewidth]{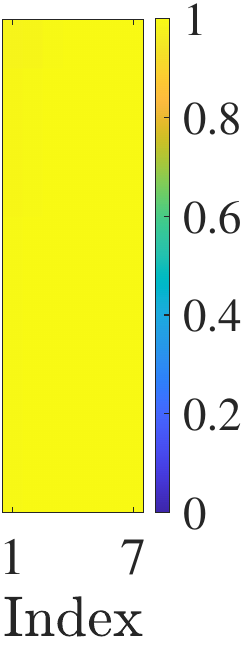}
		}
		\caption{Weight matrixes obtained from noise subspace generation model, which remove the DC and guard subcarriers and have $N-N_g-N_{dc}-1=53$ and $M-1=7$ optional eigenvectors in total for delay and Doppler noise subspace generation, respectively.}
		\label{fig:weightwith}
	\end{figure}

	Fig. \ref{fig:weightwith} shows the weight matrix obtained from noise subspace generation model and Eq. \eqref{eqn:weights}, where $\eta = 10$. Observed that during delay estimation, the model assigns higher weights to the eigenvectors corresponding to larger eigenvalues, and as the SNR increases, the range of selected eigenvectors expands to include those corresponding to smaller eigenvalues. Notably, although we encourage weight binarization through Eq. \eqref{eqn:weights}, all eigenvectors are retained in the weight matrix for delay estimation (i.e., $\omega >0$, $ \forall \omega \in \boldsymbol{\omega}_b$). This suggests that retaining all eigenvectors for constructing the noise subspace is effective, which contrasts with the conclusion presented in \cite{eigVectorSelection}. Furthermore, in Doppler estimation, all eigenvectors are selected for constructing the noise subspace (i.e., their weights $\omega =1$, $ \forall \omega \in \boldsymbol{\omega}_b$), which further explains why the performance of the enhanced model aligns with that of the traditional model, as shown in Fig. \ref{fig:Doppler_DCguard}.

	\begin{figure}
		\centering
		\includegraphics[width=0.99\linewidth]{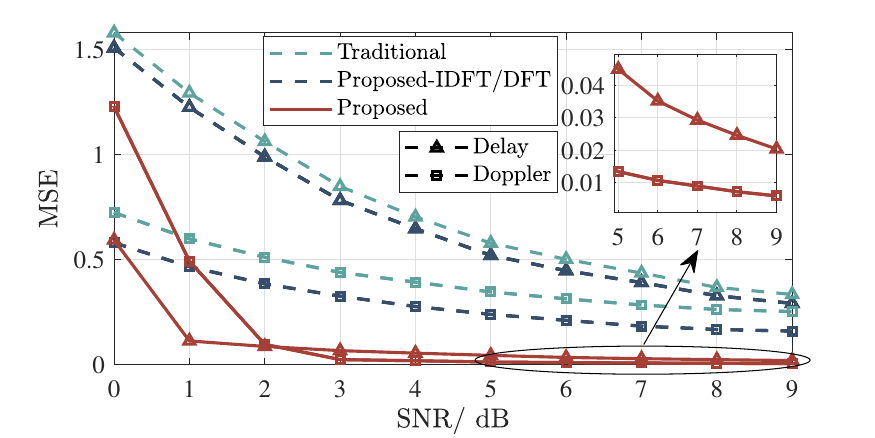}
		\caption{Parameters estimation performance with proposed noise subspace generation, peak detection and DL-IDFT/DFT models. }
		\label{fig:MSE_full}
	\end{figure}

\begin{figure}
		\centering
		\subfloat[Delay]{\label{fig:DCguarddelay_full}
			\includegraphics[width=0.76\linewidth]{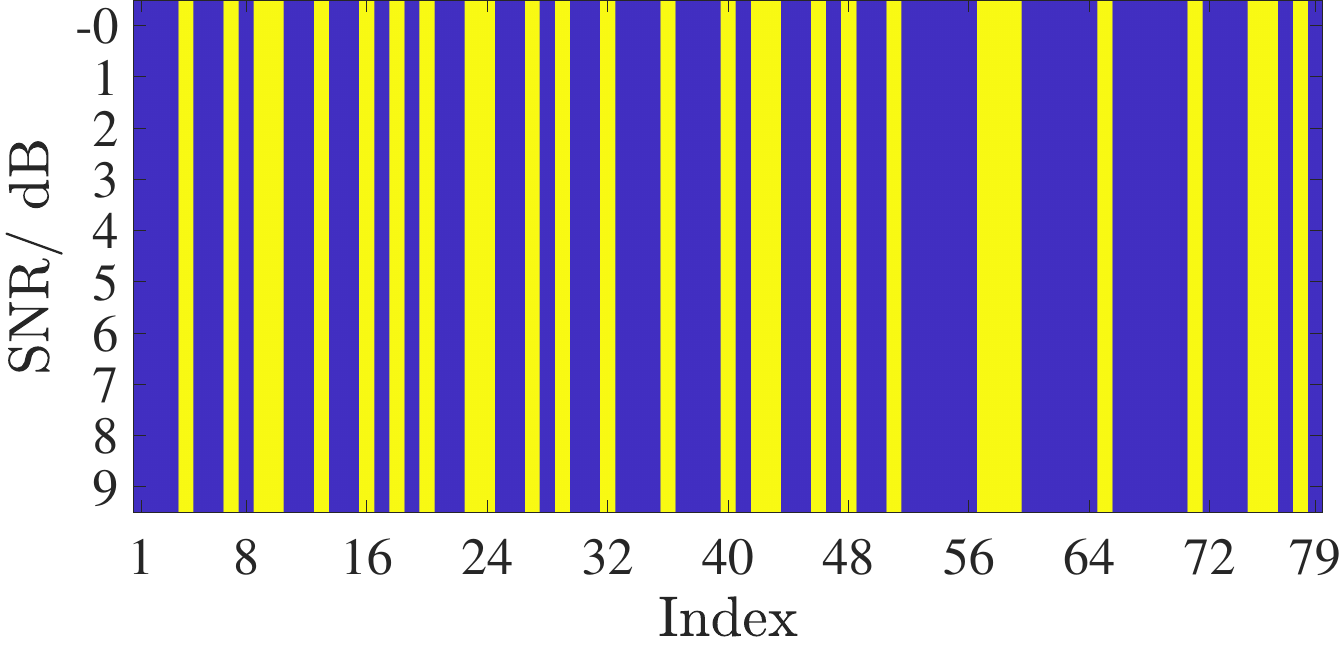}
		}
		\hspace{-2mm}
		\subfloat[Doppler]{\label{fig:DCguardDoppler_full}
			\includegraphics[width=0.145\linewidth]{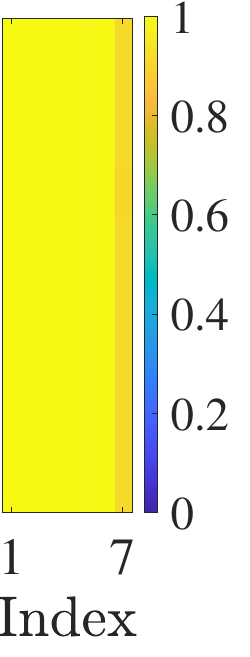}
		}
		\caption{Weight matrixes obtained from the proposed model with DL-IDFT/DFT. We have $N+N_{cp}-1=79$ and $M-1=7$ optional eigenvectors in total for delay and Doppler noise subspace generation.}
		\label{fig:weightfull}
	\end{figure}

	As a natural extension of the bedrock model, we replace the conventional IDFT/DFT in the aforementioned model with the proposed DL-IDFT/DFT, as depicted in Fig. \ref{fig:sensingpipeline}. We then comprehensively evaluated the overall performance. Although Fig. \ref{fig:Delay_DCguard} clearly illustrates the benefits of eliminating the DC and guard subcarriers, we retain the full frame as input to leverage the robust adaptability of the deep learning model. For reference, we also plot the performance from Fig. \ref{fig:MSEdelaydoppler}, where label \textit{Proposed-IDFT/DFT} corresponds to the proposed noise subspace generation and peak detection model with IDFT/DFT. We evaluate the performance only within the training SNR range of $0\sim9$ dB, and the results are presented in Fig. \ref{fig:MSE_full} and Fig. \ref{fig:weightfull}.

	Fig. \ref{fig:MSE_full} demonstrates that the parameters estimation performance compared with \textit{Proposed-IDFT/DFT} and \textit{Traditional} models. With the combined effect of the proposed DL-IDFT/DFT, noise subspace generation and peak detection models, the performance of delay and Doppler parameter estimation has been improved by an order of magnitude. Additionally, Fig. \ref{fig:weightfull} presents the eigenvector weight matrix under the proposed bedrock model. Different from the weight matrix based on IDFT/DFT, the weight matrix derived from DL-IDFT/DFT remains invariant to changes in SNR and exhibits binarization (i.e., values of either 0 or 1). This feature allows us to remove the noise subspace generation model after training, simplifying the process to a fixed selection strategy.

	\begin{figure}
		\centering
		\includegraphics[width=0.99\linewidth]{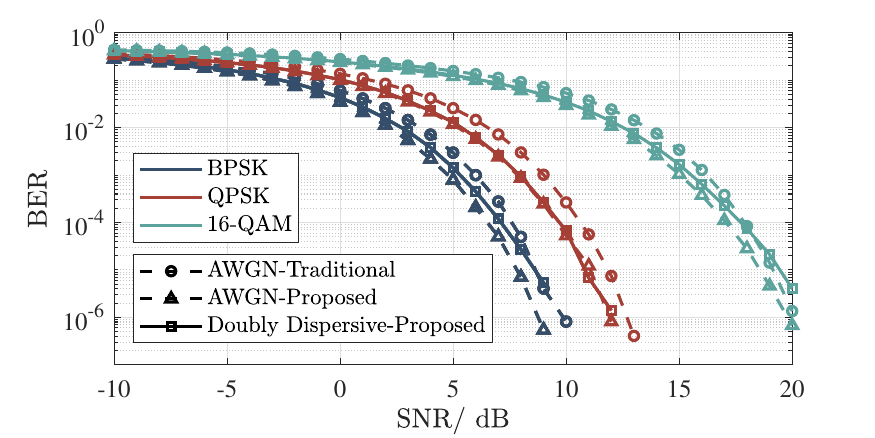}
		\caption{Sensing-aided communication performance with pre-equalization.}
		\label{fig:exp_final}
	\end{figure}
	\subsection{Performance for Sensing-aided Communications}
	As the final simulation, we integrate the pre-equalization model proposed in Section \ref{sec:sensing_aidedComm} to evaluate the performance of sensing-aided communication. Notably, we share and freeze the parameters of the communication transceiver trained under the AWGN channel, including constellation mapping, DL-IDFT/DFT, and data extraction models. We then train only the pre-equalization model at the transmitter. Fig. \ref{fig:weightfull} illustrates the performance of sensing-aided communication under this configuration. It can be observed that by leveraging the sensing results, i.e., delay and Doppler, and applying pre-equalization at the transmitter, the doubly dispersive channel at the receiver is effectively restored to an AWGN channel. This allows the E2E model to achieve performance comparable to that of the AWGN channel, even in the doubly dispersive environment. This demonstrates that pre-equalization at the transmitter can significantly reduce the signal processing complexity at the receiver. Moreover, for varying channel conditions (corresponding to different delays and dopplers), satisfactory communication performance can be achieved simply by adjusting or retraining the transmitters pre-equalization model. This provides a critical foundation for the design and deployment of future deep learning-based communication systems.

	\section{Conclusions}\label{sec:conclusion}
	In this paper, we have proposed a mathematically-based modular model, termed the bedrock model, designed for communication and sensing systems. The modular nature of the model facilitates its seamless integration into existing communication and sensing networks, enhancing generalization and transferability while improving performance. In addition, we have provided a detailed explanation of the motivations behind the design of the bedrock models, along with performance guarantees for their application in sensing and communication systems. In communication systems operating over additive white Gaussian noise (AWGN) channels, the proposed model yields a performance improvement of approximately 2 dB at low signal-to-noise ratios (SNRs), and a gain of about 0.8 dB at high SNRs. Furthermore, the deep learning-based inverse discrete Fourier transform (DL-IDFT)/DFT model exhibits notable transferability, achieving performance comparable to fully end-to-end (E2E) training by updating only the data extraction model at the receiver. In sensing applications, the integration of the bedrock models significantly enhances the traditional multiple signal classification (MUSIC) algorithm, reducing sensing errors by an order of magnitude relative to conventional methods. Additionally, a sensing-aided communication pre-equalization model within the bedrock framework is proposed. By utilizing communication models and parameters trained in AWGN channels, this model achieves near-optimal performance in doubly dispersive channels. These capabilities collectively highlight the practical feasibility of deploying the bedrock models in real-world systems.

	Several open challenges related to the proposed bedrock models remain, which will be addressed in future work due to page limitations. One promising direction is the training of the DL-IDFT/DFT model with orthogonality constraints, which could effectively mitigate intersymbol interference (ISI) while providing mathematical simplifications for easier analysis. Additionally, multi-parameter sensing tasks and iterative sensing schemes warrant further investigation, as they may offer significant advantages for practical implementation. Moreover, the performance of the proposed pre-equalization scheme under conditions involving errors in sensing parameters remains uncertain. Developing strategies to ensure robustness against such errors will be essential to align with the practical requirements of deployment.
	\bibliography{Reference}
\end{document}